\documentclass{article}

\usepackage{amsmath}
\usepackage{arxiv}
\usepackage[utf8]{inputenc} 
\usepackage[T1]{fontenc}    
\usepackage{hyperref}       
\usepackage{url}            
\usepackage{booktabs}       
\usepackage{amsfonts}       
\usepackage{nicefrac}       
\usepackage{microtype}      
\usepackage{mathtools}
\usepackage{url}
\usepackage{float}
\usepackage{placeins}
\usepackage{setspace}
\usepackage{adjustbox}
\usepackage{lineno}
\usepackage{natbib}

\makeatletter
\newif\ifnobrackets
\renewcommand\@cite[2]{\ifnobrackets\else[\fi{#1\if@tempswa , #2\fi}\ifnobrackets\else]\fi\nobracketsfalse}

\makeatother

\title{Approaching geoscientific inverse problems with vector-to-image domain transfer networks}

\author{
   Eric Laloy\thanks{Corresponding Author} \\
   Belgian Nuclear research Centre, SCK$\bullet$CEN\\
   \texttt{elaloy@sckcen.be} \\
   \And
   Niklas Linde \\
   University of Lausanne\\
   \And
   Diederik Jacques \\
   Belgian Nuclear research Centre, SCK$\bullet$CEN\\
}

\begin{document}
\maketitle

\begin{abstract}
We present \textit{vec2pix}, a deep neural network designed to predict categorical or continuous 2D subsurface property fields from one-dimensional measurement data (e.g., time series), thereby offering a new approach to solve inverse problems. The performance of the method is investigated through two types of synthetic inverse problems: (a) a crosshole ground penetrating radar (GPR) tomography experiment with GPR travel times being used to infer a 2D velocity field, and (2) a multi-well pumping experiment within an unconfined aquifer with time series of transient hydraulic heads being used to retrieve a 2D hydraulic conductivity field. For each type of problem, both a multi-Gaussian and a binary channelized subsurface domain with long-range connectivity are considered. Using a training set of 20,000 examples (implying as many forward model evaluations), the method is found to recover a 2D model that is in much closer agreement with the true model than the closest training model in the forward-simulated data space. Further testing with smaller training sample sizes shows only a moderate reduction in performance when using 5000 training examples only. Even if the recovered models are visually close to the true ones, the data misfits associated with their forward responses are generally larger than the noise level used to contaminate the true data. If finding a model that fits the data noise level is required, then \textit{vec2pix}-based inversion models can be used as initial inputs for more traditional multiple-point statistics inversion. Uncertainty of the inverse solution is partially assessed using deep ensembles, in which the network is trained repeatedly with random initialization. Overall, this study advances understanding of how to use deep learning to infer subsurface models from indirect measurement data.
\end{abstract}


\section{Introduction}
\label{intro}

Deep learning \citep[DL, see, e.g., the textbook by][]{Goodfellow2016} is currently having a profound impact on the Earth sciences \citep{Reichstein2019, Sun-Scanlon2019, Shen2018}. Important advances have been made for clustering and classification tasks \citep[e.g.,][]{Zhang2016}, forward proxy modeling \citep{Zhu-Zabaras2018} and learning tailor-made model encodings of complex geological priors into low-dimensional latent variables for geostatistical inversion \citep{Laloy2017,Laloy2018} and simulation \citep{Mosser2017, Laloy2018,Chan-Elsheikh2019} purposes. In remote sensing and geophysics, significant emphasis has been placed on how to turn low resolution images into high-resolution images using concepts of super resolution \citep[][]{Wang2019}. Lately, researchers in active seismics have started to approach inversion by transferring reflection data represented as images into geological images \citep{Araya-Polo2018,Mosser2018,Yang-Ma2019}. However, most geoscientific data do not lend themselves to a spatial representation that is visually similar to the type of final model that is sought. One example is hydrological time-series (pressure, temperature and concentration) measured at one or several locations that are only indirectly related to the underlying hydraulic conductivity field through a non-linear function. For the 2D-to-2D image transfer case, DL architectures have been proposed within the influential \textit{pix2pix} (and follow-up \textit{cycleGAN}) image-to-image translation framework \citep{Isola2016,Zhu2017}. Using a deep neural network (DNN) to turn geoscientific data vectors such as time-series into 2D or 3D subsurface models is a challenging task because, in contrast to the \textit{pix2pix} image-to-image translation framework, there is no low-level information shared between the two considered domains. Independently of our work, \citet{Earp-Curtis2020} recently proposed a 2D-to-2D DNN for travel time tomography that does not require common low-level information to be present, which, in principle, makes it amenable to 1D-to-2D domain transfer. In this contribution, we propose a 1D-to-2D network that takes one or multiple time series or other data represented in a data vector and map them into a subsurface model. This implies that we bypass conventional inversion by instead learning a mapping between 1D measurement data and a corresponding 2D subsurface model. Our presented examples focus on inferring ground-penetrating radar (GPR) velocity and hydraulic conductivity for given 2-D channelized and multi-Gaussian prior models, but the potential of the approach is much wider than this. To assess uncertainty in the inverse solutions, we use the recently developed deep ensembles approach \citep{Fort2019}. The results produced by our so-called \textit{vec2pix} algorithm demonstrate the feasibility of 1D-to-2D transfer, thereby allowing for many possible applications in hydrology, geophysics and Earth system science.

The remainder of this paper is organized as follows. Section \ref{related_work} summarizes related work and how it differs from our method. Section \ref{methods} describes our proposed domain transfer network and its training, together with the considered inverse problems. This is followed by section \ref{results} that presents our domain transfer inversion results. In section \ref{discussion}, we discuss our main findings and outline current limitations and possible future developments. Finally, section \ref{conclusion} provides a conclusion. 

\section{Related Work}
\label{related_work}

Inversion using image-to-image domain transfer networks has been proposed in the context of 2D seismic inversion \citep{Araya-Polo2018,Mosser2018,Yang-Ma2019}. In subsurface hydrology, the study by \citet{Sun2018} is a first step towards inverting steady-state groundwater flow data with image-to-image domain translation. Both \citet{Mosser2018} and \citet{Sun2018} added loss functions to the \textit{cycleGAN} network by \citet{Zhu2017} to promote reconstruction of paired images. The works listed so far require that the two considered 2D domains share some low-level information. As written above, independently of our work \citet{Earp-Curtis2020} proposed a 2D-to-2D transfer network for inversion of 2D travel time tomography data which does not have that requirement. In this study, we explicitly cast the problem within a vector-to-image transform framework. Our network is however conceptually similar to that of \citet{Earp-Curtis2020} in the sense that the 1D input processed by our network gets projected in a 2D space at some point (see section \ref{network}). The main differences between our work and the study by \citet{Earp-Curtis2020} are as follows. First our work is rooted within an informative prior framework aiming at obtaining solutions with a high degree of prescribed geological realism \citep{Linde2015}. While \citet{Earp-Curtis2020} use a completely uncorrelated prior parameter space, we consider the common case in hydrogeology and hydrogeophysics where prior information on the considered subsurface structure is available under the form of a geologically-based prior model. Compared to \citet{Earp-Curtis2020}, this allows us to work with significantly higher-dimensional output model domains and, perhaps more importantly, with much less training samples for learning the weights and biases of our network. Indeed, we consider (at most) 20,000 training samples and model sizes of $128 \times 64$ and $160 \times256$. In contrast, \citet{Earp-Curtis2020} consider small  $8 \times 8$ and  $16 \times 16$ model domains, and use as many as 2.5 million training samples (i.e., 125 times more). This has a high impact on computational feasibility as obtaining one training sample requires one forward model evaluation. Second, \citet{Earp-Curtis2020} focus on 2D travel time tomography only while we also consider a rather nonlinear transient groundwater flow problem. Lastly, we use a different neural network architecture.

\section{Methods}
\label{methods}
\subsection{Vector-to-image transfer network}
\label{methods_network}

Let us denote by $Y$ the measurement data (vector) domain, and by $X$ the model (2D subsurface property field) domain. Our model consists of the mapping function, $G_{YX}$:
	\begin{equation}
	\label{eq1}
	G_{YX} : \mathbb{R}^Y \rightarrow \mathbb{R}^X.
	\end{equation}
The $G_{YX}$ operator predicts the model $\tilde{\textbf{x}}$ corresponding to the measurement data vector $\textbf{y}$ it is fed with, $\tilde{\textbf{x}} = G_{YX}\left(\textbf{y}\right)$. At training time, $G_{YX}$ is learned using a l$_{\rm 1}$ reconstruction loss: 
	\begin{equation}
	\label{eq2}
	\min_{G_{YX}}\left\{\mathcal{L}_{\rm rec}\left(G_{YX}, \textbf{y}, \textbf{x} \right)\right\} = \underset{\textbf{x}\sim p_{\textbf{x}}}{\mathbb{E}}\left[||\textbf{x} - G_{YX}\left(\textbf{y}\right)||_1\right].
	\end{equation}

\subsection{Implementation}
\label{network}

\begin{figure}[H]
	\noindent\hspace{-0cm}\includegraphics[width=40pc]{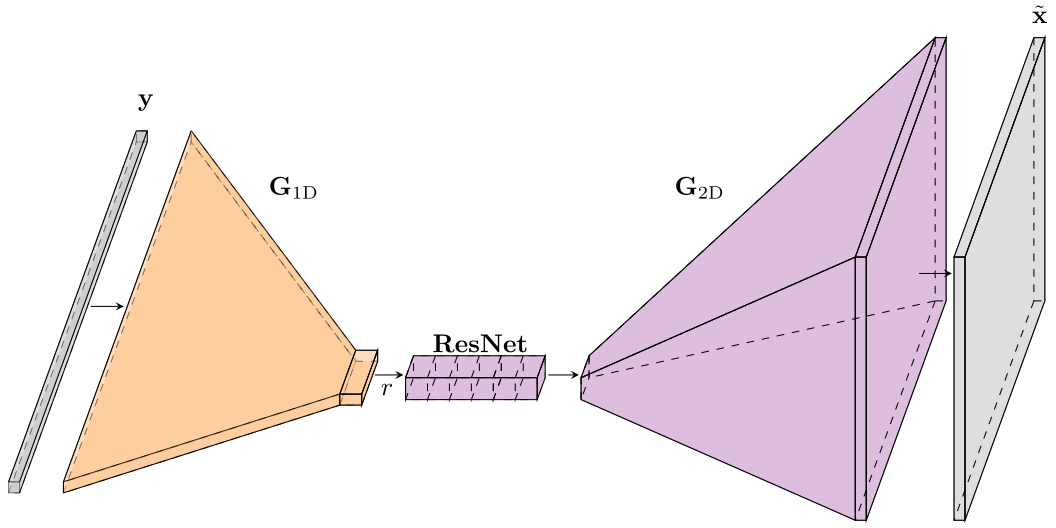}
	\caption{Simplified illustration of the \textit{vec2pix} architecture. The gray signifies the input data vector (left) and output image or model (right). The 1D input data are standardized such that they have a zero-mean and unit variance. The output models are in $\left[-\textbf{1},\textbf{1}\right]$. The orange and violet colors denote the 1D and 2D parts of the network, respectively. The sketch is not to scale and the depths of the convolutions are either represented by a single extra unit length in the horizontal direction (central orange rectangle and Resnet blocks) or are not represented at all (orange and violet trapezoids). More specifically, $\textbf{y}$ and $\tilde{\textbf{x}}$ are the input vector and reconstructed model, respectively, $\textbf{G}_{\rm 1D}$ denotes the series of 1D convolutions with increasing depths, $r$ represents the 1D-to-2D reshape operation, $\textbf{G}_{\rm 2D}$ signifies the series of 2D transposed convolutions and final 2D convolution with decreasing depths, and $\textbf{ResNet}$ refers to the ensemble of ``ResNet" residual blocks.} 
	\label{fig1}
\end{figure}

As stated above, the main methodological difference between our network architecture and most of those used previously within the 2D-to-2D transform paradigm \citep{Isola2016,Zhu2017,Mosser2018,Sun2018} is that our code processes a 1D (input) vector to output a 2D array. Our generator network architecture is based on \citet{Zhu2017} and follows the state-of-the-art in computer vision. To make our \textit{vec2pix} generator suitable to the 1D-to-2D ($G_{YX}$) domain transfer, we first project the input data vector onto an increasingly larger number of lower-dimensional representations (or latent spaces or manifolds) using a series of 1D convolutions with increasing number of channels (or filters, see Figure \ref{fig1} and Appendix A). Then we apply a reshape operation to convert the final 1D representations into 2D representations before (i) further processing this information through a series of so-called ``ResNet" residual blocks \citep{He2016} and (ii) projecting the derived latent spaces into increasingly larger-dimensional representations while reducing their numbers, until the final 2D model is produced. This is achieved by using a combination of 2D transposed convolutions and a final 2D convolution (Figure \ref{fig1}). The key step of going from a 1D to a 2D domain therefore consists in the simple yet practical reshaping operation. Our generator is detailed in \ref{appendix}. Note that projecting the input data onto an increasingly large number of low-dimensional representations allows our network to learn many different features from the input data. If not all of the different low-dimensional representations are needed to perform the mapping between the considered domains, then during training the network is expected to learn the relevant representations only. 

\begin{figure}[H]
	\noindent\hspace{1cm}\includegraphics[width=35pc]{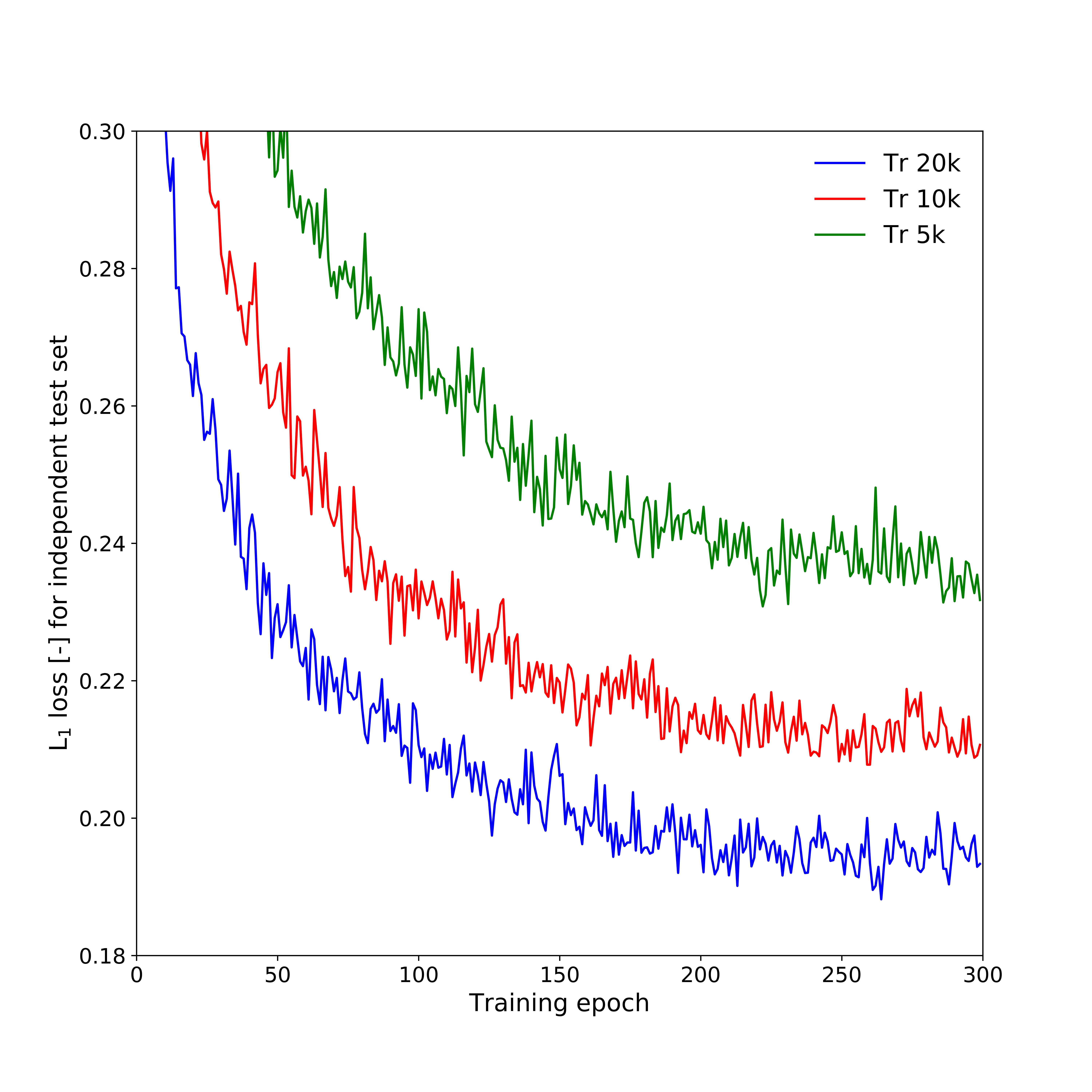}
	\caption{Convergence of the used (mean l$_{\rm 1}$) reconstruction loss for an independent test set of 100 samples and Case study 4: transient pressure data and binary channelized domain. The Tr 5k, Tr 10k and Tr 20k labels denote training with 5000 training samples, 10,000 training samples and 20,000 training samples, respectively.} 
	\label{fig2}
\end{figure}

The Adam optimization solver \citep{Kingma-Ba2015} was used for training. We used a learning rate of 0.00001 for the multi-Gaussian case and a learning rate of 0.0002 for the categorical case (see section \ref{inv_prob} for details about these two cases), and values of 0.5 and 0.999 for the $\beta_1$ and $\beta_2$ momentum parameters. For the multi-Gaussian case, the \textit{vec2pix} model realizations were post-processed with a median filter (see section \ref{inv_prob} for details). Unless stated otherwise, the number of epochs used in training is 200 for the GPR case studies (see section \ref{inv_gpr}) and 300 for the flow case studies (see section \ref{inv_flow}), and the batch size is 25. For every experiment, we first used 20,000 examples of $\textbf{x}_i$ - $\textbf{y}_i$ pairs. To study the sensitivity of our results to the training set size, we also considered training with 5,000 and 10,000 training examples for every case study (see section \ref{size_tr}). We used an additional small validation set of 100 pairs (unseen by the training algorithm) to monitor the evolution of the loss function during training (see Figure \ref{fig2}). The indices of the input-output training pairs are shuffled at the beginning of every epoch to help the gradient descent escape local minima. Furthermore, to make training robust to the noise in the data, that is, to account for the data measurement error during training, each true data vector used for training was corrupted with a new Gaussian white noise realization prior to the next epoch. With respect to performance evaluation, an independent test set made of 1000 examples was used to assess the performance of the proposed approach. Hence, inversion performance is assessed by evaluating how well each of those 1000 test models are recovered when the trained $G_{YX}$ transformer is fed with the corresponding noise-contaminated data.

\FloatBarrier

\subsection{Synthetic Inverse Problems}
\label{inv_prob}
To test \textit{vec2pix}, we consider both crosshole ground penetrating radar (GPR) data and transient pressure data acquired during pumping. As for prior geologic models, we consider two common cases: a 2D multi-Gaussian prior and a 2D binary channelized aquifer prior. Regarding the latter, the DeeSse (DS) MPS algorithm \citep{Mariethoz2010a} was used to generate the training and test models from the channelized aquifer training image proposed by \citet{Zahner2016}. To produce the multi-Gaussian realizations for training and test purposes, we used the circulant embedding method \citep{Dietrich-Newsam1997}. For the multi-Gaussian case, the \textit{vec2pix} predictions were postprocessed by application of a median filter with a kernel size of either 3 (GPR case) or 5 (hydraulic case) pixels in each spatial direction. No postprocessing was applied to the \textit{vec2pix} predictions for the categorical case, except for thresholding before running the forward solver.

\subsubsection{Crosshole GPR data}
\label{inv_gpr}
Crosshole GPR imaging uses a transmitter antenna to emit a high-frequency electromagnetic wave at a location in one borehole and a receiver antenna to record the arriving energy at a location in another borehole. The considered synthetic measurement data are first-arrival travel times for several transmitter and receiver locations. These data contain information about the GPR velocity distribution between the boreholes. The GPR velocity primarily depends on dielectric permittivity, which is strongly influenced by water content and, consequently, porosity in saturated media \citep{Roth1990}. The considered model domain is of size 128 $\times$ 64 with a cell size of 0.1 m, and our setup consists of two vertical boreholes that are located 6.4 m apart placed at the left and right hand sides of the domain. Sources (left) and receivers (right) are located between 0.5 and 12.5 m depth with 0.5 m spacing (Figures \ref{fig3}a and \ref{fig3}c), leading to a total dataset of $\textbf{y} = \textbf{d}_{\rm GPR}$ of 625 travel times. The forward nonlinear ray-based response is simulated by the pyGIMLi toolbox \citep{Rucker2017} using the Dijkstra method. The measurement error used to corrupt the data is a zero-mean uncorrelated Gaussian with a standard deviation of 0.5 ns, which is typical for high-quality GPR field data.

For the binary channelized aquifer case, the channel and background facies are assigned velocities of 0.06 m ns$^{-1}$ and 0.08 m ns$^{-1}$, respectively (Figure \ref{fig3}c). For the multi-Gaussian case, a zero-mean anisotropic Gaussian covariance model with a variance (sill) of 0.5, integral scales in the horizontal and vertical directions of 2 m (20 pixels) and 4 m (40 pixels), respectively, and anisotropy angle of 60$^{\circ}$ was selected. The model realizations, were then scaled in $\left[-1,1\right]$ using the minimum and maximum pixel values over the 20,000 training models before the following relationship was used to convert a scaled model, $\textbf{x}$, into a velocity model, $\textbf{x}_{\rm VEL} = 0.06 + 0.02\left(\textbf{1}-\textbf{x}\right)$ m/ns (Figure \ref{fig3}a). For illustrative purposes, the simulated data vectors corresponding to the models depicted in Figures \ref{fig3}a and \ref{fig3}c are shown in Figures \ref{fig4}a and \ref{fig4}c.

\subsubsection{Transient pumping data}
\label{inv_flow}
\begin{figure}[hbt!]
	\noindent\hspace{-3cm}\includegraphics[width=45pc]{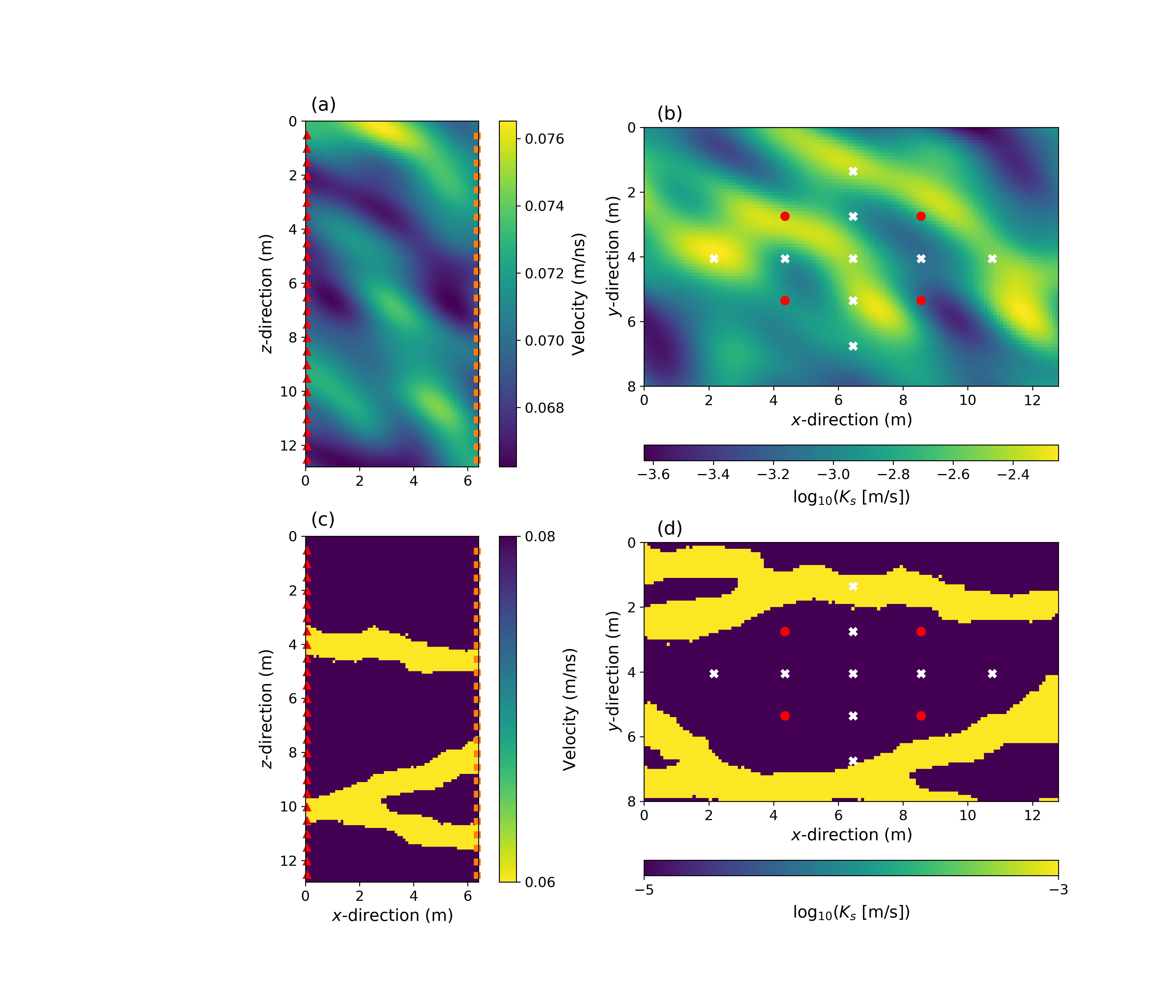}
	\caption{Four considered synthetic case studies. The GPR tomography case with (a) multi-Gaussian and (c) channelized bimodal surface structures with red triangles and orange squares representing the GPR source and receiver positions, respectively. The transient pumping cases with (b) multi-Gaussian and (d) channelized bimodal surface structures and red dots and white crosses representing the pumping/observation and observation-only wells, respectively. The models displayed in subfigures (a-d) are randomly chosen from the 20,000 training models considered for each of the four examples.} 
	\label{fig3}
\end{figure}

\begin{figure}[hbt!]
	\noindent\hspace{-1cm}\includegraphics[width=45pc]{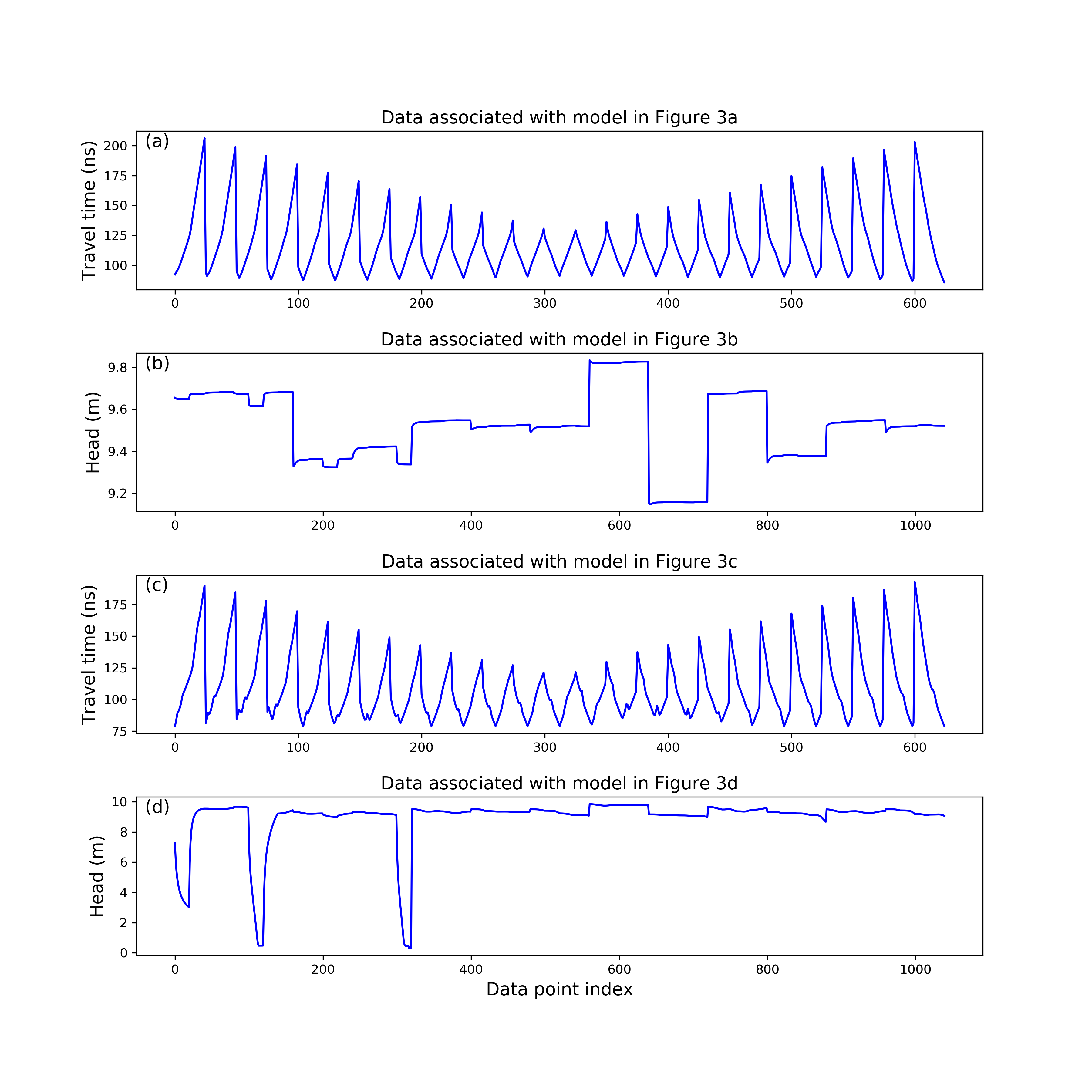}
	\caption{Simulated data corresponding to the training models depicted in Figure \ref{fig3}.} 
	\label{fig4}
\end{figure}

Our second type of data consists of transient piezometric heads induced by pumping. The 80 $\times$ 128 aquifer domain lies in the $x-y$ plane with a grid cell size of 1 m and a thickness of 10 m. For the binary channelized aquifer case, channel and matrix materials (see Figure \ref{fig3}d) are assigned hydraulic conductivity values,  $K_s$, of 1 $\times$ 10$^{-3}$ m/s and 1 $\times$ 10$^{-5}$ m/s, respectively. For the multi-Gaussian case, the same geostatistical parameters as for the GPR setup are used for $\log_{10}\left(K_s\right)$, except that the mean is -3 and the variance 0.1. The assumed specific storage and specific yield of the aquifer are 0.0003 m$^{-1}$ and 0.3 (-), respectively. The MODFLOW-NWT \citep{Niswonger2011} code is used to simulate unconfined transient groundwater flow with no-flow boundaries at the upper and lower sides and a lateral head gradient of 0.01 (-) with water flowing in the $x$-direction. Four wells are sequentially extracting water for 20 days at a rate of 0.001 m$^3$/s (red dots in Figures \ref{fig3}b and d). The measurement data were formed by acquiring daily simulated heads in the four pumping wells (red dots in Figures \ref{fig3}b and d) and nine measuring wells (white crosses in Figures \ref{fig3}b and \ref{fig3}d) during the 80 days simulation period. The synthetic measurement data comprises $\textbf{y} = \textbf{d}_{\rm Flow}$ of 1040 heads. The standard deviation of the measurement error used to contaminate these data with a Gaussian white noise is set to 0.01 m. Figures \ref{fig4}b and \ref{fig4}d display the concatenated data vectors corresponding to the models depicted in Figures \ref{fig3}b and \ref{fig3}d. 

\FloatBarrier

\subsection{Uncertainty Quantification}
\label{methods_uq}

Most high-dimensional inverse problems are under-determined, implying that the mapping from noise-contaminated data to a model is non-unique. For this reason, it is important to explore different mappings such that uncertainty in the inverse solution can be assessed. Recent work on understanding the loss landscape of high-dimensional deep networks \citep[][]{Fort-Jastrzebski2019,Fort2019} has shown that building deep ensembles by training the network multiple times using random initialization of weights and biases works well empirically, and is currently the most viable strategy for exploring multi-modal landscapes. In particular, it has been found to outperform Monte Carlo dropout and various subspace sampling strategies, which often drastically underestimate uncertainty \citep[][]{Fort2019}. Here we adopt such an ensemble framework to investigate predictive uncertainty, using a small ensemble of five trained models. That said, we stress that uncertainty quantification in the context of deep learning is still largely an unresolved problem. 

\section{Results}
\label{results}

For each of the four considered case-studies, we investigate the performance of $G_{YX}$ based on the independent $i=1,\cdots, 1000$ test pairs of model, $\textbf{x}_i$,  and data, $\textbf{y}_i$. For each test model, $\textbf{x}_i$, the root-mean-square error (RMSE) between the associated data $\textbf{y}_i$ and the $j=1,\cdots,20000$ training data vectors $\textbf{y}_j$ is computed and the minimum RMSE over the resulting 20,000 values is retained as the smallest distance in data space between the considered test model and the training set. On this basis, we specifically compare the true and predicted model for cases where:
\begin{enumerate}
	\item The true model is taken as the most different test model from the set of training models in the data space.
	\item The true model is taken as the second most different test model from the set of training models in the data space.
	\item A representative model of the test set is chosen and this procedure is repeated four times.
\end{enumerate}
Cases 1 and 2 serve to highlight the capacity of \textit{vec2pix} to generalize for cases that are distinctively different from the training data. This leads to six cases where differences between true models and those predicted by \textit{vec2pix} are scrutinized. For each case we perform two predictions (\#1 and \#2) based on two different noise realizations used to corrupt the true measurement data. This is done to assess the impact of the measurement data noise realization on prediction accuracy, which should be limited because of our robust training strategy. Thus, the smaller the differences between these two model predictions the better. These two predictions are not be confounded with our main uncertainty quantification estimates which, a stated earlier, are based on deep ensembles (see section \ref{pred_unc}). Furthermore, the complete distribution of 1000 RMSEs between the test data and the data simulated by feeding the forward solver with the models predicted by $G_{YX}$ for these test data is also considered. In addition, two similarity indices between true and generated \textit{vec2pix} models are computed for the 1000 test examples: the ${\rm l}_1$ norm, and the widely-used structural similarity index (SSIM) \citep{Wang2004}
	\begin{equation}
	\label{eq3}
	{\rm SSIM}\left( \textbf{u}, \textbf{v}\right) = \frac{2\mu_{\textbf{u}}\mu_{\textbf{v}}+c_12\sigma_{\textbf{uv}}+c_2}{\mu_{\textbf{u}}^2+\mu_{\textbf{v}}^2+c_1\sigma^2_{\textbf{u}}+\sigma^2_{\textbf{v}}+c_2},
	\end{equation}
where $\textbf{u}$ and $\textbf{v}$ denote two $N_p \times N_p$ windows subsampled from $\textbf{x}$ and $ \tilde{\textbf{x}}$, respectively, $\mu$ and $\sigma^2$ are the mean and variance of $\textbf{u}$ and $\textbf{v}$, $\sigma_{\textbf{uv}}$ represents the covariance between $\textbf{u}$ and $\textbf{v}$, and $c_1 = 0.01$ and $c_2 = 0.03$ are two small constants \citep{Wang2004}. Averaged over all  $\textbf{u}$ and $\textbf{v}$ sliding windows, the mean SSIM ranges from -1 to 1, with 1 implying that the two compared images are identical. Similarly to \citet{Sun2018} and \citet{Earp-Curtis2020}, we set $N_p = 7$.

\subsection{Case study 1: crosshole GPR data and multi-Gaussian model domain}
\label{results:case1}
\begin{figure}[hbt!]
	\noindent\hspace{-4cm}\includegraphics[width=60pc]{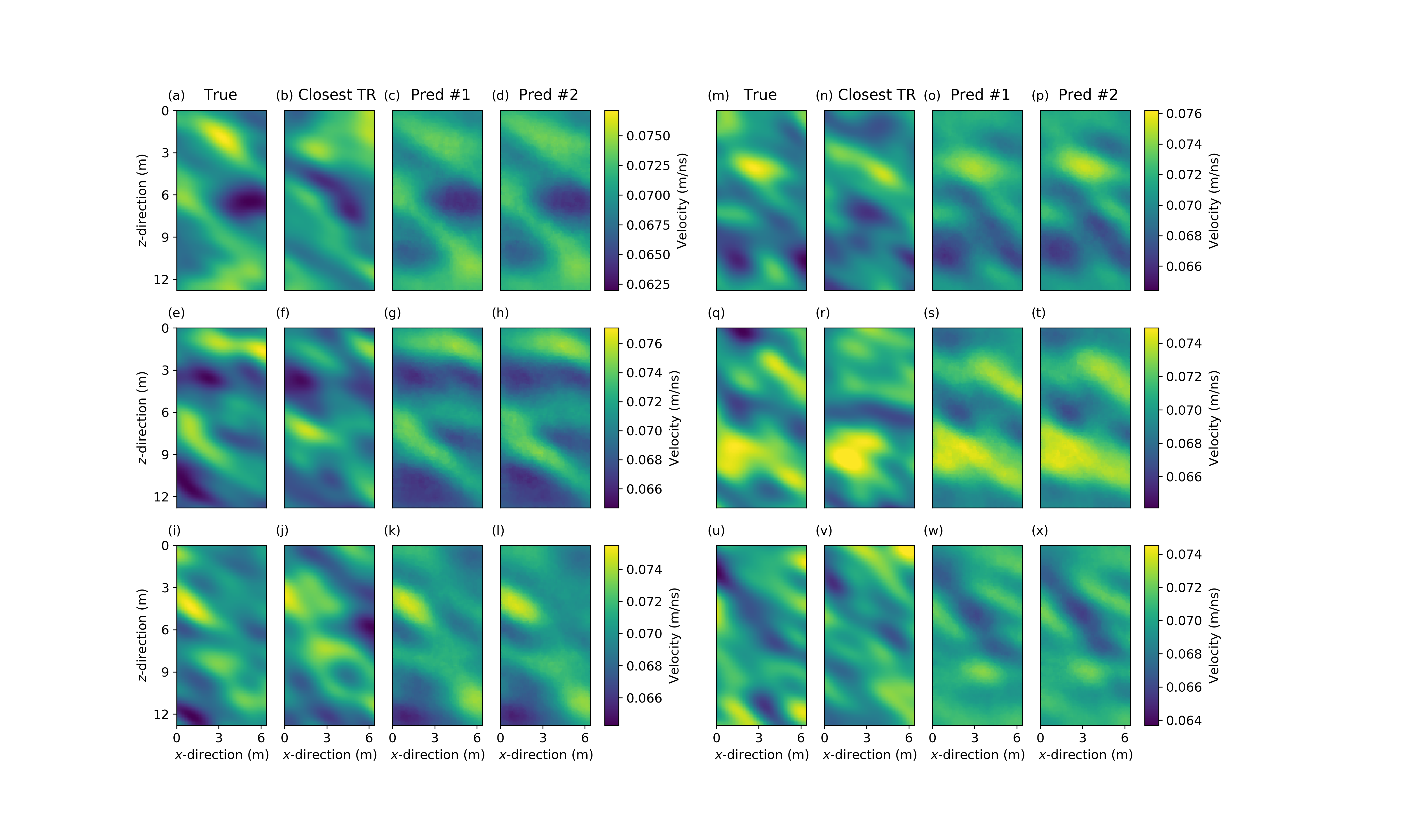}
	\caption{\small{Results for crosshole GPR data in a multi-Gaussian domain. (a-d) The true model is the most different test model from the 20,000 training models in data space: (a) true model, (b) closest training model in data space, (c, d) predicted models from two different noise realizations (Pred \#1 and Pred \#2). The same plotting style is adapted for cases (e-h) where the true model is the second most different test model in data space, and (i-l), (m-p), (q-t) and (m-x) for four representative test models. Table \ref{table1} lists the prediction quality statistics associated with the models displayed in the (a - x) subfigures.}} 
	\label{fig5}
\end{figure}

The \textit{vec2pix} results for the GPR first-arrival travel time tomography within a multi-Gaussian domain are presented in Figure \ref{fig5} and Table \ref{table1} for the six selected true models, while Table \ref{table2} lists the corresponding performance statistics for the 1000 test examples. The produced \textit{vec2pix} models always induce a lower data misfit and are more similar to the true test models than the corresponding closest training models in data space (Tables \ref{table1} and \ref{table2}). For instance, the \textit{vec2pix} models display a two times smaller ${\rm l}_1$-norm than the closest training models in data space (Table \ref{table1}). Also, the SSIMs of the \textit{vec2pix} models are 10\% to 30\% larger than those of the closest training models in data space (Tables \ref{table1} and \ref{table2}). Using 20,000 training examples, it is consequently a better option to train and use \textit{vec2pix} to invert the ``measurement" data than to simply pick up the training model with the best corresponding data fit. This shows that \textit{vec2pix} can generalize. The data RMSE of the forward-simulated \textit{vec2pix} realizations are globally in the 0.5 ns - 0.9 ns range with a median of 0.58 ns (Table \ref{table2}), which is reasonably close to the ``true" noise level of 0.5 ns used to contaminate the data (``true" measurement error). Overall, as compared to using the closest training model, \textit{vec2pix} allows for a reduction in data RMSE by a factor of two for the considered multi-Gaussian problem (Table \ref{table1}). However, the \textit{vec2pix} models are a bit too smooth.

\begin{table}[hbt!]
	\caption{\small{Statistics of the results obtained for crosshole GPR data in a multi-Gaussian domain. The naming convention is the same and the letters (a-x) refer to the models in Figure \ref{fig4}. RMSE$_{\rm data}$ denotes the RMSE in data space, ${\rm l}_1$ refers to the ${\rm l}_1$-norm and SSIM to the structural similarity index. The ${\rm l}_1$-norm is calculated in terms of velocity (m/ns) while the SSIM is computed in the rescaled $\left[0,1\right]$ domain. Closest TR means the closest training model in data space and Predicted \#1 and Predicted \#2 signify predicted models from two different noise realizations.}}
	\begin{adjustbox}{center}
		\begin{tabular}{cccc}%
			\hline
			True model & RMSE$_{\rm data}$ (ns) & ${\rm l}_1$ (m/ns) & SSIM (-)\\
			\hline
			(a) True & 0.5 & 0 & 1 \\
			(b) Closest TR & 1.73 & 18.92 & 0.72 \\
			(c) Predicted \#1 & 0.72 & 8.07 & 0.92 \\
			(d) Predicted \#2 & 0.73 & 8.10 & 0.92 \\
			
			(e) True & 0.5 & 0 & 1 \\
			(f) Closest TR & 1.44 & 15.35 & 0.77 \\
			(g) Predicted \#1 & 0.63 & 6.26 & 0.93 \\
			(h) Predicted \#2 & 0.68 & 6.60 & 0.92 \\
			
			(i) True & 0.5 & 0 & 1 \\
			(j) Closest TR & 0.98 & 13.31 & 0.79 \\
			(k) Predicted \#1 & 0.56 & 6.26 & 0.93 \\
			(l) Predicted \#2 & 0.59 & 5.89 & 0.94 \\
			
			(m) True & 0.5 & 0 & 1 \\
			(n) Closest TR & 1.01 & 12.37 & 0.82 \\
			(o) Predicted \#1 & 0.60 & 7.20 & 0.93 \\
			(p) Predicted \#2 & 0.63 & 7.23 & 0.92 \\
			
			(q) True & 0.5 & 0 & 1 \\
			(r) Closest TR & 1.16 & 13.17 & 0.78 \\
			(s) Predicted \#1 & 0.58 & 5.89 & 0.93 \\
			(t) Predicted \#2 & 0.61 & 6.65 & 0.92 \\
			
			(u) True & 0.5 & 0 & 1 \\
			(v) Closest TR & 1.05 & 12.65 & 0.81 \\
			(w) Predicted \#1 & 0.63 & 7.83 & 0.90 \\
			(x) Predicted \#2 & 0.67 & 7.55 & 0.91 \\
			
			\hline
		\end{tabular}
	\end{adjustbox}
	\label{table1}
\end{table}
\begin{table}[hbt!]
	\caption{\small{Statistics for the case of crosshole GPR data in a multi-Gaussian domain when considering the 1000 independent test models. The comparison is made between the predicted model (Predicted \#1) and the closest training model in data space (Closest TR) within the training set of 20,000 examples using the RMSE in data space ( RMSE$_{\rm data}$), the ${\rm l}_1$-norm (${\rm l}_1$) and the structural similarity index (SSIM). For each metric, the minimum (Min), median (Median) and maximum (Max) values are reported together with the 10$^{\rm th}$, 25$^{\rm th}$, 75$^{\rm th}$ and 90$^{\rm th}$ percentiles (P10, P25, P75, P90), respectively. The TR size variable signifies the number of training examples used to train \textit{vec2pix}.}}
	\begin{adjustbox}{center}
		\begin{tabular}{ccccccccc}%
			\hline
			Model & TR size & Min & P10 & P25 & Median & P75 & P90 & Max\\
			\hline
			& & & & & & & & \\
			\multicolumn{9}{c}{RMSE$_{\rm data}$ (ns)} \\
			Closest TR & 20,000 & 0.77 & 0.92 & 0.97 & 1.03 & 1.10 & 1.16 & 1.73\\
			Predicted & 20,000 & 0.50 & 0.55 & 0.56 & 0.58 & 0.61 & 0.63 & 0.92 \\
			Predicted & 10,000 &  0.53 & 0.59 & 0.61 & 0.64 & 0.68 & 0.73 & 1.12 \\
			Predicted & 5,000 &  0.56 & 0.63 & 0.67 & 0.81 & 1.08 & 1.39 & 2.24  \\
			\multicolumn{9}{c}{${\rm l}_1$ (m/ns)} \\
			Closest TR & 20,000 & 6.92 & 10.57 & 11.41 & 12.36 & 13.56 & 14.56 & 19.27 \\
			Predicted & 20,000 & 3.68 & 5.19 & 5.80 & 6.50 & 7.35 & 8.28 & 12.16 \\
			Predicted & 10,000 & 4.29 & 5.77 & 6.47 & 7.24 & 8.16 & 9.05 & 13.48  \\
			Predicted & 5,000 & 4.20 & 6.48 & 7.11 & 7.97 & 8.83 & 10.00 & 13.14  \\
			\multicolumn{9}{c}{SSIM (-)}\\
			Closest TR & 20,000 & 0.66 & 0.75 & 0.78 & 0.80 & 0.83 & 0.84 & 0.91 \\
			Predicted & 20,000 & 0.87 & 0.91 & 0.92 & 0.93 & 0.94 & 0.95 & 0.96 \\
			Predicted & 10,000 &  0.84 & 0.89 & 0.91 & 0.92 & 0.93 & 0.94 & 0.96  \\
			Predicted & 5,000 &  0.84 & 0.88 & 0.89 & 0.91 & 0.92 & 0.93 & 0.95  \\
			\hline
		\end{tabular}
	\end{adjustbox}
	\label{table2}
\end{table}
\FloatBarrier
\subsection{Case study 2: crosshole GPR data and binary channelized domain}
\label{results:case2}

\begin{figure}[hbt!]
	\noindent\hspace{-4cm}\includegraphics[width=60pc]{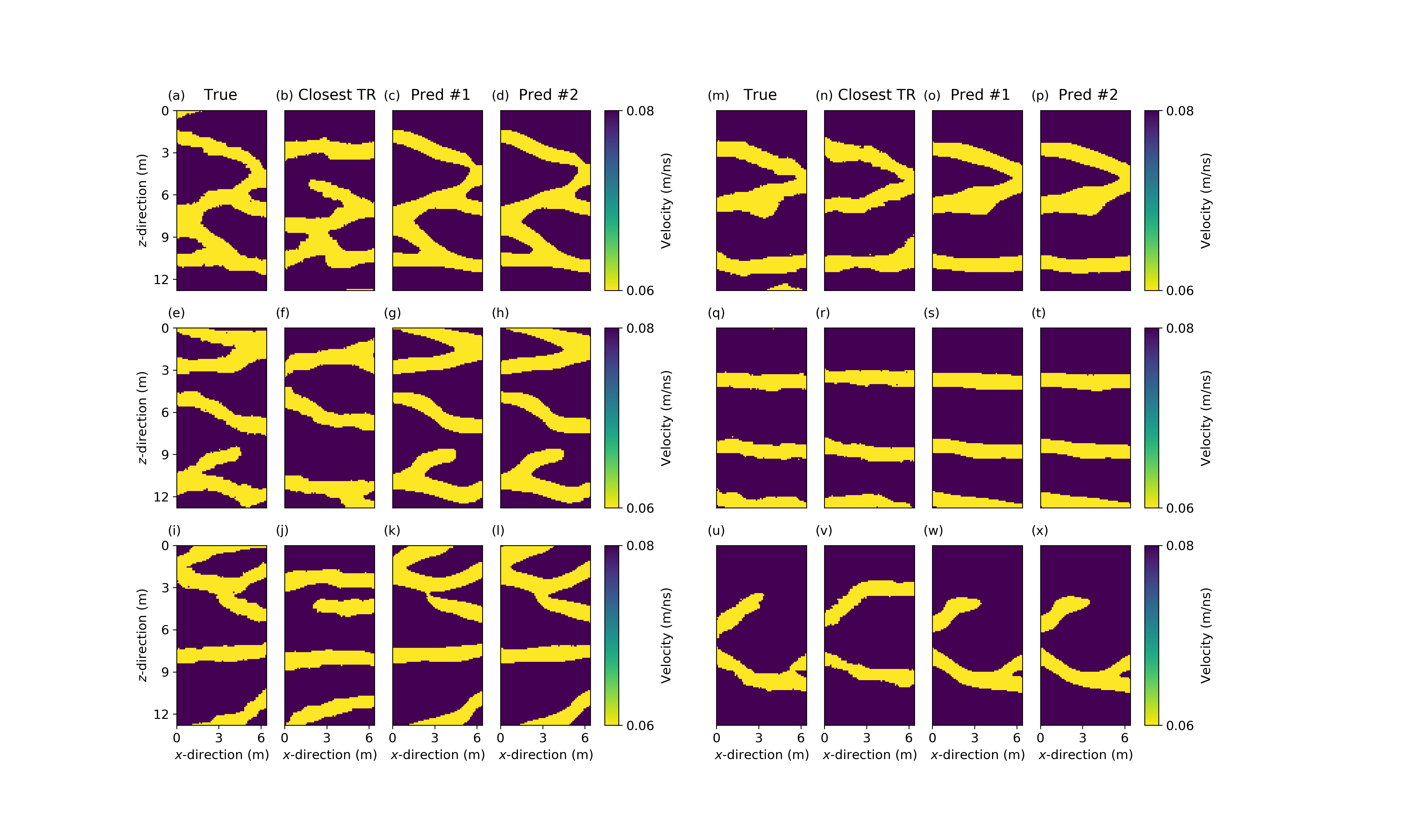}
	\caption{\small{Results for crosshole GPR data in a binary channelized domain. (a-d) The true model is the most different test model from the 20,000 training models in data space: (a) true model, (b) closest training model in data space, (c, d) predicted models from two different noise realizations (Pred \#1 and Pred \#2). The same plotting style is adapted for cases (e-h) where the true model is the second most different test model in data space, and (i-l), (m-p), (q-t) and (m-x) for four representative test models. Table \ref{table3} lists the prediction quality statistics associated with the models displayed in the (a - x) subfigures.}} 
	\label{fig6}
\end{figure}

Results for travel time tomography within a binary channelized domain are displayed in Figure \ref{fig6}, Table \ref{table3} and Table \ref{table4}. These results are in line with those obtained for the multi-Gaussian case: the predicted test models show lower data RMSE, lower $\rm{l}_1$ and larger SSIM statistics than the closest training models in data space. Also, the predicted test models look visually close to their true counterparts. With a median data RMSE of 0.87 ns and min and max data RMSEs of 0.58 ns and 1.98 ns (Table \ref{table4}), the predicted test models induce data RMSE values that are significantly larger than the ``true" noise level of 0.5 ns. Nevertheless, \textit{vec2pix} permits reduction in data RMSE of a factor two to three compared to the closest corresponding training model (Table \ref{table3}). The associated SSIM indices are smaller than for the multi-Gaussian case: the P10 and median SSIM values are 0.72 and 0.81 (Table \ref{table4}) against 0.91 and 0.93 for the multi-Gaussian case (Table \ref{table2}). Globally, despite leading to larger data RMSEs compared to the prescribed noise level of 0.5 ns, the \textit{vec2pix} models are similar to the true ones (Figure \ref{fig6} and Tables \ref{table3}).

\begin{table}[hbt!]
	\caption{\small{Statistics of the results obtained for crosshole GPR data in a binary channelized domain. The naming convention is the same and the letters (a-x) refer to the models in Figure \ref{fig5}. RMSE$_{\rm data}$ denotes the RMSE in data space, ${\rm l}_1$ refers to the ${\rm l}_1$-norm and SSIM to the structural similarity index. The ${\rm l}_1$-norm is calculated in terms of velocity (m/ns) while the SSIM is computed in the rescaled $\left[0,1\right]$ domain. Closest TR means the closest training model in data space and Predicted \#1 and Predicted \#2 signify predicted models from two different noise realizations.}}
	\begin{adjustbox}{center}
		\begin{tabular}{cccc}%
			\hline
			True model & RMSE$_{\rm data}$ (ns) & ${\rm l}_1$ (m/ns)& SSIM (-)\\
			\hline
			(a) True & 0.5 & 0 & 1 \\
			(b) Closest TR & 3.50 & 45.36 & 0.40 \\
			(c) Predicted \#1 & 1.05 & 11.04 & 0.71 \\
			(d) Predicted \#2 & 1.06 & 10.98 & 0.72 \\
			
			(e) True & 0.5 & 0 & 1 \\
			(f) Closest TR & 3.06 & 42.42 & 0.41 \\
			(g) Predicted \#1 & 1.31 & 14.42 & 0.67 \\
			(h) Predicted \#2 & 1.26 & 15.20 & 0.66 \\
			
			(i) True & 0.5 & 0 & 1 \\
			(j) Closest TR & 2.46 & 49.28 & 0.39 \\
			(k) Predicted \#1 & 1.24 & 9.90 & 0.78 \\
			(l) Predicted \#2 & 0.90 & 10.50 & 0.77 \\
			
			(m) True & 0.5 & 0 & 1 \\
			(n) Closest TR & 2.14 & 20.06 & 0.60 \\
			(o) Predicted \#1 & 0.99 & 10.70 & 0.74 \\
			(p) Predicted \#2 & 0.98 & 9.46 & 0.77 \\
			
			(q) True & 0.5 & 0 & 1 \\
			(r) Closest TR & 1.20 & 14.92 & 0.70 \\
			(s) Predicted \#1 & 0.66 & 4.90 & 0.87 \\
			(t) Predicted \#2 & 0.61 & 4.52 & 0.88 \\
			
			(u) True & 0.5 & 0 & 1 \\
			(v) Closest TR & 2.54 & 24.66 & 0.64 \\
			(w) Predicted \#1 & 0.96 & 5.24 & 0.86 \\
			(x) Predicted \#2 & 0.90 & 5.88 & 0.84 \\
			
			\hline
		\end{tabular}
	\end{adjustbox}
	\label{table3}
\end{table}

\begin{table}[hbt!]
	\caption{\small{Statistics for the case of crosshole GPR data in a binary channelized domain when considering the 1000 independent test models. The comparison is made between the predicted model (Predicted \#1) and the closest training model in data space (Closest TR) within the training set of 20,000 examples using the RMSE in data space ( RMSE$_{\rm data}$), the ${\rm l}_1$-norm (${\rm l}_1$) and the structural similarity index (SSIM). For each metric, the minimum (Min), median (Median) and maximum (Max) values are reported together with the 10$^{\rm th}$, 25$^{\rm th}$, 75$^{\rm th}$ and 90$^{\rm th}$ percentiles (P10, P25, P75, P90), respectively. The TR size variable signifies the number of training examples used to train \textit{vec2pix}.}}
	\begin{adjustbox}{center}
		\begin{tabular}{ccccccccc}%
			\hline
			Model & TR size & Min & P10 & P25 & Median & P75 & P90 & Max\\
			\hline
			& & & & & & & & \\
			\multicolumn{9}{c}{RMSE$_{\rm data}$ (ns)} \\
			Closest TR & 20,000 & 0.77 & 1.31 & 1.53 & 1.78 & 2.09 & 2.35 & 3.50\\
			Predicted & 20,000 & 0.58 & 0.70 & 0.77 & 0.87 & 1.00 & 1.13 & 1.98\\
			Predicted & 10,000 & 0.56 & 0.79 & 0.89 & 1.02 & 1.17 & 1.40 & 2.36  \\
			Predicted & 5,000 &  0.58 & 0.83 & 0.94 & 1.11 & 1.32 & 1.57 & 2.59  \\
			\multicolumn{9}{c}{${\rm l}_1$ (m/ns)} \\
			Closest TR & 20,000 & 4.82 & 13.88 & 17.59 & 22.42 & 28.97 & 35.29 & 56.74 \\
			Predicted & 20,000 & 2.06 & 4.46 & 6.04 & 7.76 & 10.03 & 12.48 & 24.88\\
			Predicted & 10,000 &  3.06 & 5.80 & 7.48 & 9.70 & 12.53 & 15.16 & 25.20  \\
			Predicted & 5,000 &  2.78 & 6.18 & 8.36 & 10.89 & 14.60 & 17.46 & 27.32  \\
			\multicolumn{9}{c}{SSIM (-)}\\
			Closest TR & 20,000 & 0.32 & 0.50 & 0.56 & 0.63 & 0.70 & 0.75 & 0.90 \\
			Predicted & 20,000 & 0.59 & 0.72 & 0.77 & 0.81 & 0.85 & 0.88 & 0.94 \\
			Predicted & 10,000 &  0.57 & 0.69 & 0.73 & 0.78 & 0.82 & 0.86 & 0.92  \\
			Predicted & 5,000 &  0.54 & 0.66 & 0.71 & 0.77 & 0.81 & 0.85 & 0.93  \\
			\hline
		\end{tabular}
	\end{adjustbox}
	\label{table4}
\end{table}
\FloatBarrier

\subsection{Case study 3: transient pressure data and multi-Gaussian domain}
\label{results:case3}

\begin{figure}[hbt!]
	\noindent\hspace{-2cm}\includegraphics[width=50pc]{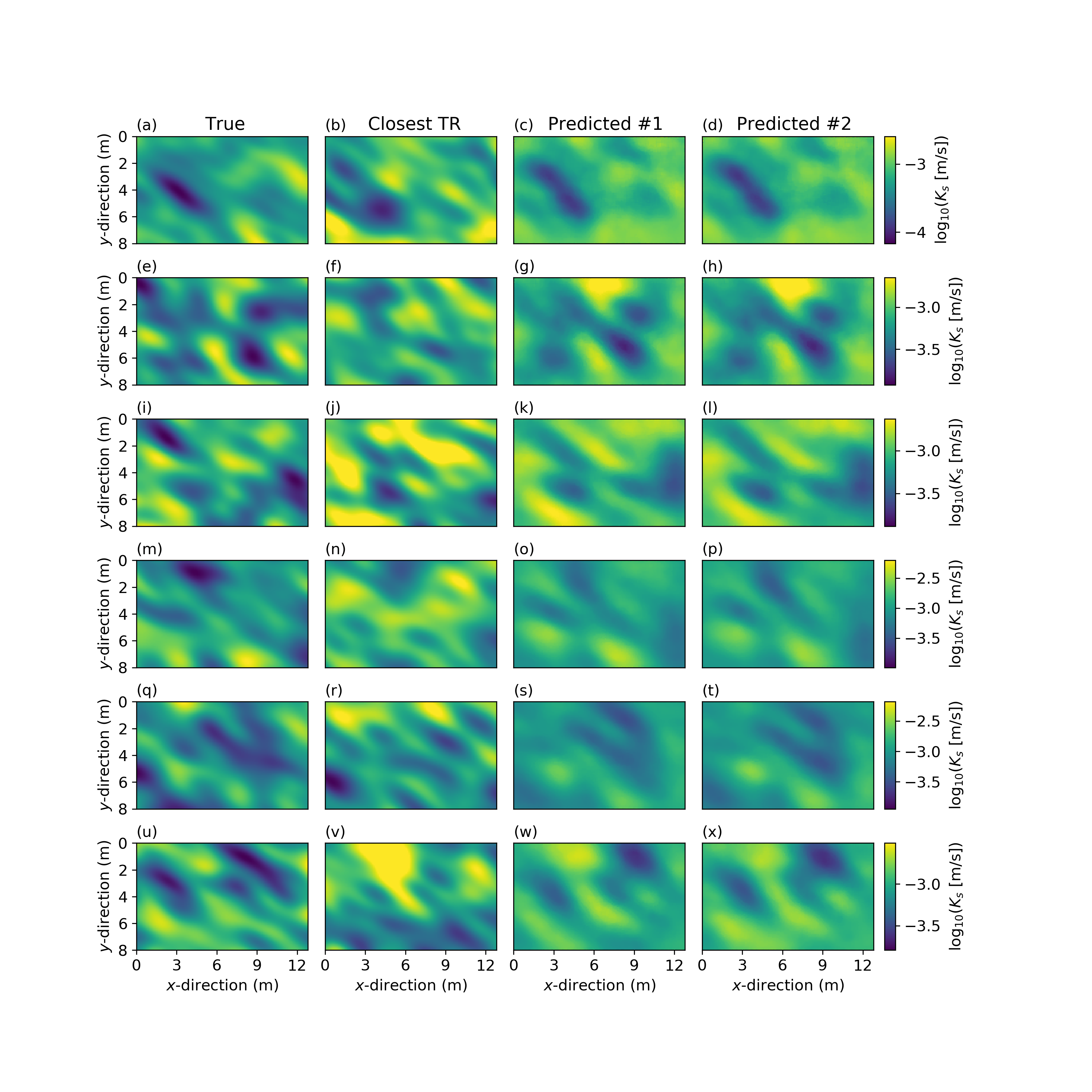}
	\caption{\small{Results for transient subsurface pressure data in a multi-Gaussian domain. (a-d) The true model is the most different test model from the 20,000 training models in data space: (a) true model, (b) closest training model in data space, (c, d) predicted models from two different noise realizations (Pred \#1 and Pred \#2). The same plotting style is adapted for cases (e-h) where the true model is the second most different test model in data space, and (i-l), (m-p), (q-t) and (m-x) for four representative test models. Table \ref{table5} lists the prediction quality statistics associated with the models displayed in the (a - x) subfigures.}} 
	\label{fig7}
\end{figure}

For the transient pumping experiment within a multi-Gaussian domain, the \textit{vec2pix} models are visually close to the true ones (Figure \ref{fig7}), even if they appear slightly too smooth. The RMSEs in data space produced by the \textit{vec2pix} models are overall similar to those produced by the closest training models (Tables \ref{table5} and \ref{table6}), and are mostly distributed in the 0.02 m - 0.03 m range that is to be compared with the ``true" noise level of 0.01 m. However, the model reconstruction statistics, ${\rm l}_1$-norm and SSIM, are substantially better for the \textit{vec2pix} models than for the closest training models in data space (Tables \ref{table5} and \ref{table6}). Indeed, the \textit{vec2pix} models display 40\% to 60\% smaller ${\rm l}_1$-norms and 15\% to 25\% larger SSIMs.

\begin{table}[hbt!]
	\caption{\small{Statistics of the results obtained for transient subsurface pressure data in a multi-Gaussian domain. The naming convention is the same and the letters (a-x) refer to the models in Figure \ref{fig6}. RMSE$_{\rm data}$ denotes the RMSE in data space, ${\rm l}_1$ refers to the ${\rm l}_1$-norm and SSIM to the structural similarity index. The ${\rm l}_1$-norm is calculated in terms of $\log_{10}K_s$ (-) while the SSIM is computed in the rescaled $\left[0,1\right]$ domain. Closest TR means the closest training model in data space and Predicted \#1 and Predicted \#2 signify predicted models from two different noise realizations.}}
	\begin{adjustbox}{center}
		\begin{tabular}{cccc}%
			\hline
			True model & RMSE$_{\rm data}$ (m) & ${\rm l}_1$ (m) & SSIM (-)\\
			\hline
			(a) True & 0.010 & 0 & 1 \\
			(b) Closest TR & 0.060 & 2807 & 0.79 \\
			(c) Predicted \#1 & 0.023 & 1837 & 0.90 \\
			(d) Predicted \#2 & 0.021 & 1788 & 0.90 \\
			
			(e) True & 0.010 & 0 & 1 \\
			(f) Closest TR & 0.049 & 2886 & 0.77 \\
			(g) Predicted \#1 & 0.054 & 1559 & 0.92 \\
			(h) Predicted \#2 & 0.041 & 1629 & 0.91 \\
			
			(i) True & 0.010 & 0 & 1 \\
			(j) Closest TR & 0.027 & 2848 & 0.75 \\
			(k) Predicted \#1 & 0.018 & 1612 & 0.93 \\
			(l) Predicted \#2 & 0.021 & 1503 & 0.93 \\
			
			(m) True & 0.010 & 0 & 1 \\
			(n) Closest TR & 0.022 & 3575 & 0.76 \\
			(o) Predicted \#1 & 0.017 & 1913 & 0.89 \\
			(p) Predicted \#2 & 0.023 & 1820 & 0.90 \\
			
			(q) True & 0.010 & 0 & 1 \\
			(r) Closest TR & 0.024 & 3122 & 0.74 \\
			(s) Predicted \#1 & 0.017 & 1954 & 0.88 \\
			(t) Predicted \#2 & 0.019 & 2048 & 0.88 \\
			
			(u) True & 0.010 & 0 & 1 \\
			(v) Closest TR & 0.020 & 3644 & 0.76 \\
			(w) Predicted \#1 & 0.016 & 1524 & 0.91 \\
			(x) Predicted \#2 & 0.013 & 1442 & 0.91 \\
			
			\hline
		\end{tabular}
	\end{adjustbox}
	\label{table5}
\end{table}

\begin{table}[hbt!]
	\caption{\small{Statistics for the case of transient subsurface pressure data in a multi-Gaussian domain when considering the 1000 independent test models. The comparison is made between the predicted model (Predicted \#1) and the closest training model in data space (Closest TR) within the training set of 20,000 examples using the RMSE in data space ( RMSE$_{\rm data}$), the ${\rm l}_1$-norm (${\rm l}_1$) and the structural similarity index (SSIM). For each metric, the minimum (Min), median (Median) and maximum (Max) values are reported together with the 10$^{\rm th}$, 25$^{\rm th}$, 75$^{\rm th}$ and 90$^{\rm th}$ percentiles (P10, P25, P75, P90), respectively. The TR size variable signifies the number of training examples used to train \textit{vec2pix}.}}
	\begin{adjustbox}{center}
		\begin{tabular}{ccccccccc}%
			\hline
			Model & TR size & Min & P10 & P25 & Median & P75 & P90 & Max\\
			\hline
			& & & & & & & & \\
			\multicolumn{9}{c}{RMSE$_{\rm data}$ (m)} \\
			Closest TR & 20,000 & 0.015 & 0.019 & 0.021 & 0.023 & 0.026 & 0.029 & 0.061\\
			Predicted & 20,000 & 0.011 & 0.016 & 0.019 & 0.024 & 0.031 & 0.039 & 0.070 \\
			Predicted & 10,000 &  0.011 & 0.017 & 0.020 & 0.025 & 0.034 & 0.046 & 0.091  \\
			Predicted & 5,000 &  0.013 & 0.021 & 0.026 & 0.033 & 0.043 & 0.055 & 0.096  \\
			\multicolumn{9}{c}{${\rm l}_1$ (m)} \\
			Closest TR & 20,000 & 1819 & 2408 & 2616 & 2848 & 3088 & 3321 & 3984 \\
			Predicted & 20,000 & 1273 & 1527 & 1623 & 1753 & 1894 & 2035 & 2408 \\
			Predicted & 10,000 &  1176 & 1586 & 1703 & 1832 & 1983 & 2130 & 2597 \\
			Predicted & 5,000 & 1203 & 1600 & 1734 & 1929 & 2108 & 2294 & 3055  \\
			\multicolumn{9}{c}{SSIM (-)}\\
			Closest TR & 20,000 & 0.58 & 0.71 & 0.74 & 0.76 & 0.79 & 0.81 & 0.86 \\
			Predicted & 20,000 & 0.81 & 0.86 & 0.88 & 0.89 & 0.90 & 0.92 & 0.94 \\
			Predicted & 10,000 &  0.80 & 0.86 & 0.87 & 0.88 & 0.90 & 0.91 & 0.93  \\
			Predicted & 5,000 & 0.80 & 0.85 & 0.86 & 0.88 & 0.89 & 0.90 & 0.93  \\
			\hline
		\end{tabular}
	\end{adjustbox}
	\label{table6}
\end{table}
\FloatBarrier
\subsection{Case study 4: transient pressure data and binary channelized domain}
\label{results:case4}

\begin{figure}[hbt!]
	\noindent\hspace{-2cm}\includegraphics[width=50pc]{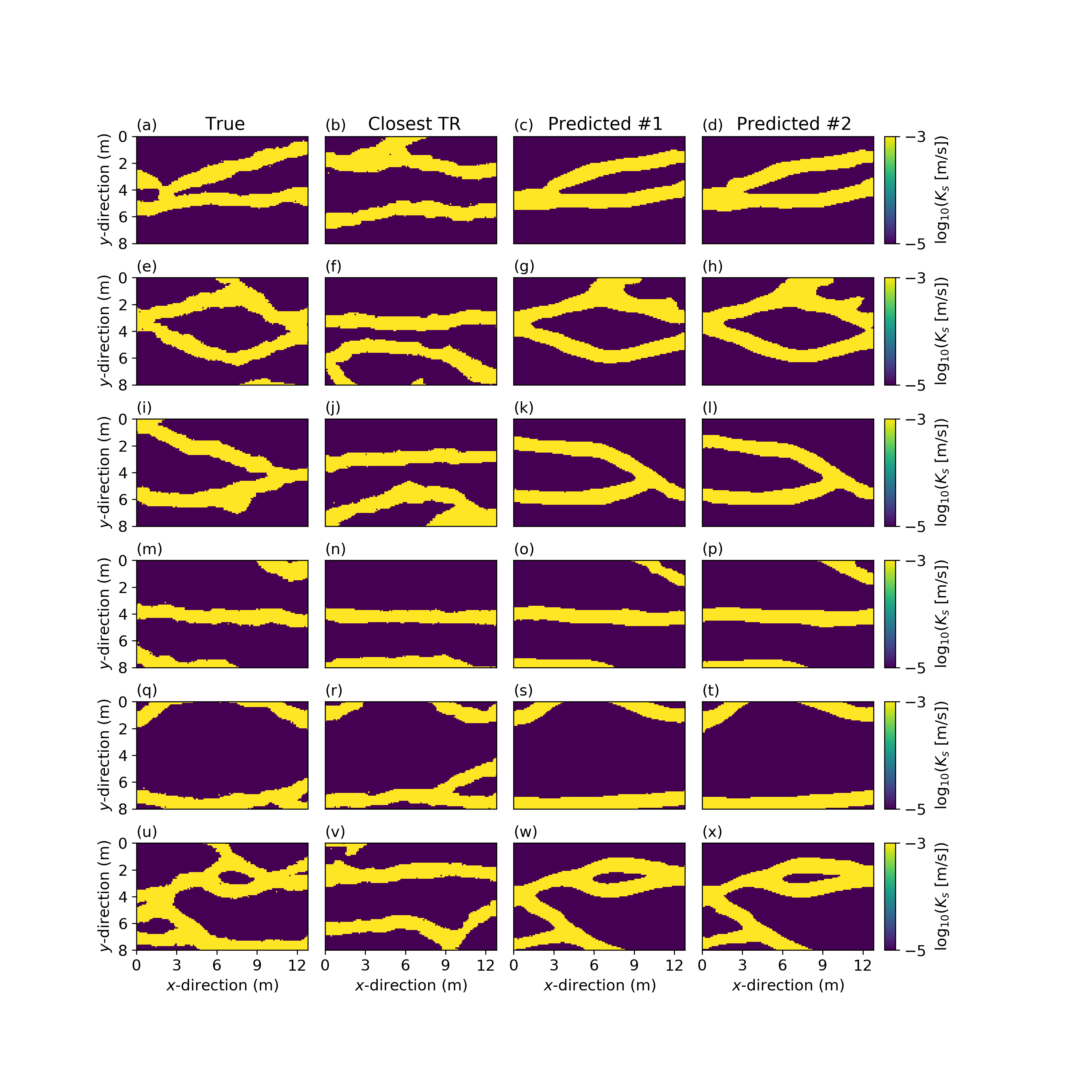}
	\caption{\small{Results for transient subsurface pressure data in a binary channelized domain. (a-d) The true model is the most different test model from the 20,000 training models in data space: (a) true model, (b) closest training model in data space, (c, d) predicted models from two different noise realizations (Pred \#1 and Pred \#2). The same plotting style is adapted for cases (e-h) where the true model is the second most different test model in data space, and (i-l), (m-p), (q-t) and (m-x) for four representative test models. Table \ref{table7} lists the prediction quality statistics associated with the models displayed in the (a - x) subfigures.}} 
	\label{fig8}
\end{figure}

The hydraulic case with a binary channelized domain is by far the most challenging as the relationship between a binary channelized model and the resulting simulated transient flow data is highly nonlinear. As a consequence, across the 20,000 training models, the signal-to-noise-ratio (SNR) defined as the ratio of the average RMSE obtained by drawing prior realizations from the training image by MPS simulation to the noise level is in the 60 - 100 range. It is seen that the \textit{vec2pix} models are in better visual agreement with the true model than the closest training models in data space (Figure \ref{fig8}). This is confirmed by two to three times smaller ${\rm l}_1$-norms and 10\% to 220\% larger SSIM indices (Tables \ref{table7} and \ref{table8}). Even if \textit{vec2pix} produces models that are of much better quality than the closest training models in data space, the resulting RMSEs in data space are often not better than those produced by the closest training models in data space. This is because a change of facies in the surroundings of a pumping well (red dots in Figure \ref{fig3}d) can dramatically affect the corresponding simulated data. 

\begin{table}[hbt!]
	\caption{\small{Statistics of the results obtained for transient subsurface pressure data in a binary channelized domain. The naming convention is the same and the letters (a-x) refer to the models in Figure \ref{fig7}. RMSE$_{\rm data}$ denotes the RMSE in data space, ${\rm l}_1$ refers to the ${\rm l}_1$-norm and SSIM to the structural similarity index. The ${\rm l}_1$-norm is calculated in terms of $\log_{10}K_s$ (-) while the SSIM is computed in the rescaled $\left[0,1\right]$ domain. Closest TR means the closest training model in data space and Predicted \#1 and Predicted \#2 signify predicted models from two different noise realizations.}}
	\begin{adjustbox}{center}
		\begin{tabular}{cccc}%
			\hline
			True model & RMSE$_{\rm data}$ (m) & ${\rm l}_1$ (m) & SSIM (-)\\
			\hline
			(a) True & 0.010 & 0 & 1 \\
			(b) Closest TR & 0.385 & 9908 & 0.21 \\
			(c) Predicted \#1 & 0.589 & 2504 & 0.67 \\
			(d) Predicted \#2 & 0.590 & 2594 & 0.67 \\
			
			(e) True & 0.010 & 0 & 1 \\
			(f) Closest TR & 0.281 & 8180 & 0.28 \\
			(g) Predicted \#1 & 0.363 & 2536 & 0.64 \\
			(h) Predicted \#2 & 0.215 & 2246 & 0.65 \\
			
			(i) True & 0.010 & 0 & 1 \\
			(j) Closest TR & 0.052 & 8028 & 0.32 \\
			(k) Predicted \#1 & 0.229 & 3068 & 0.64 \\
			(l) Predicted \#2 & 0.067 & 2950 & 0.65 \\
			
			(m) True & 0.010 & 0 & 1 \\
			(n) Closest TR & 0.037 & 2754 & 0.71 \\
			(o) Predicted \#1 & 0.021 & 1338 & 0.81 \\
			(p) Predicted \#2 & 0.033 & 1420 & 0.80 \\
			
			(q) True & 0.010 & 0 & 1 \\
			(r) Closest TR & 0.027 & 3398 & 0.68 \\
			(s) Predicted \#1 & 0.014 & 1638 & 0.80 \\
			(t) Predicted \#2 & 0.014 & 1624 & 0.80 \\
			
			(u) True & 0.010 & 0 & 1 \\
			(v) Closest TR & 0.067 & 9774 & 0.20 \\
			(w) Predicted \#1 & 0.379 & 3418 & 0.60 \\
			(x) Predicted \#2 & 0.416 & 3334 & 0.60 \\
			\hline
		\end{tabular}
	\end{adjustbox}
	\label{table7}
\end{table}

\begin{table}[hbt!]
	\caption{\small{Statistics for the case of transient subsurface pressure data in a binary channelized domain when considering the 1000 independent test models. The comparison is made between the predicted model (Predicted \#1) and the closest training model in data space (Closest TR) within the training set of 20,000 examples using the RMSE in data space ( RMSE$_{\rm data}$), the ${\rm l}_1$-norm (${\rm l}_1$) and the structural similarity index (SSIM). For each metric, the minimum (Min), median (Median) and maximum (Max) values are reported together with the 10$^{\rm th}$, 25$^{\rm th}$, 75$^{\rm th}$ and 90$^{\rm th}$ percentiles (P10, P25, P75, P90), respectively. The TR size variable signifies the number of training examples used to train \textit{vec2pix}.}}
	\begin{adjustbox}{center}
		\begin{tabular}{ccccccccc}%
			\hline
			Model & TR size & Min & P10 & P25 & Median & P75 & P90 & Max\\
			\hline
			& & & & & & & & \\
			\multicolumn{9}{c}{RMSE$_{\rm data}$ (m)} \\
			Closest TR & 20,000 & 0.011 & 0.024 & 0.030 & 0.043 & 0.064 & 0.092 & 0.385\\
			Predicted & 20,000 & 0.011 & 0.016 & 0.020 & 0.031 & 0.083 & 0.248 & 0.713\\
			Predicted & 10,000 & 0.011 & 0.018 & 0.023 & 0.039 & 0.105 & 0.276 & 8.981 \\
			Predicted & 5000 & 0.011 & 0.020 & 0.028 & 0.051 & 0.128 & 0.308 & 8.981 \\
			\multicolumn{9}{c}{${\rm l}_1$ (m)}\\
			Closest TR & 20,000 & 548 & 2335 & 3251 & 4359 & 5665 & 7049 & 11160 \\
			Predicted & 20,000 & 316 & 1064 & 1436 & 1990 & 2559 & 3222 & 6634\\
			Predicted & 10,000 &  348 & 1174 & 1608 & 2154 & 2838 & 3504 & 6692  \\
			Predicted & 5,000 &  456 & 1316 & 1796 & 2415 & 3171 & 4015 & 6750  \\
			\multicolumn{9}{c}{SSIM (-)}\\
			Closest TR & 20,000 & 0.11 & 0.37 & 0.47 & 0.55 & 0.64 & 0.72 & 0.89 \\
			Predicted & 20,000 & 0.36 & 0.62 & 0.67 & 0.73 & 0.79 & 0.84 & 0.93 \\
			Predicted & 10,000 &  0.34 & 0.59 & 0.66 & 0.71 & 0.77 & 0.82 & 0.93  \\
			Predicted & 5,000 &  0.34 & 0.56 & 0.62 & 0.70 & 0.76 & 0.81 & 0.91 \\
			\hline
		\end{tabular}
	\end{adjustbox}
	\label{table8}
\end{table}
\FloatBarrier

\subsection{Predictive Uncertainty}
\label{pred_unc}

\begin{figure}[hbt!]
	\noindent\hspace{-1.5cm}\includegraphics[width=50pc]{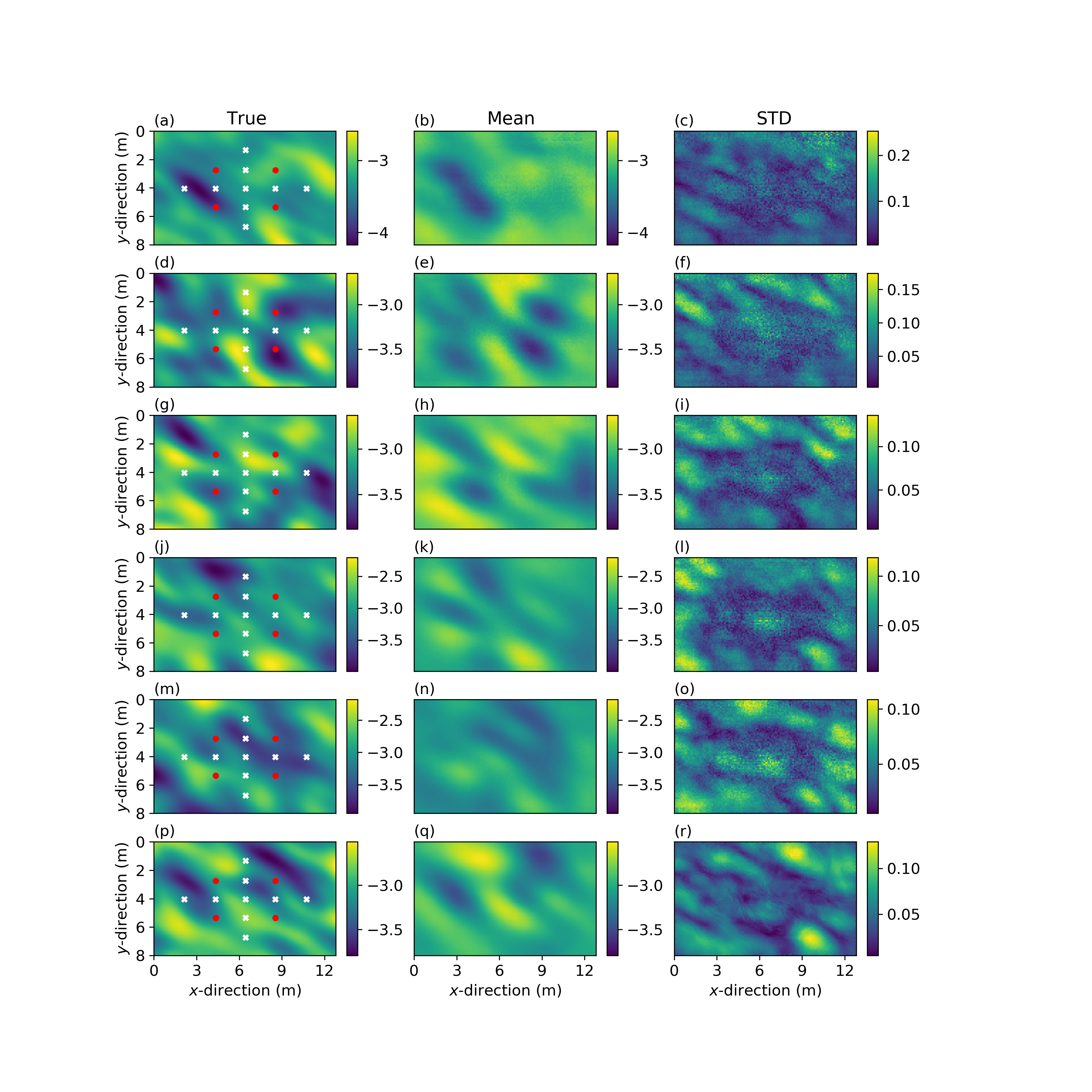}
	\caption{\small{Predictive uncertainty of inverse mapping using deep ensembles associated with the previously selected six true models for the case with transient subsurface pressure data and a multi-Gaussian domain. The True label indicates the true model while the Mean and STD labels denote the mean predicted model and its associated standard deviation map calculated over an ensemble of five members. Table \ref{table9} lists the prediction quality statistics associated with the mean model and the five ensemble models.}} 
	\label{fig9}
\end{figure}

\begin{figure}[hbt!]
	\noindent\hspace{-1.5cm}\includegraphics[width=50pc]{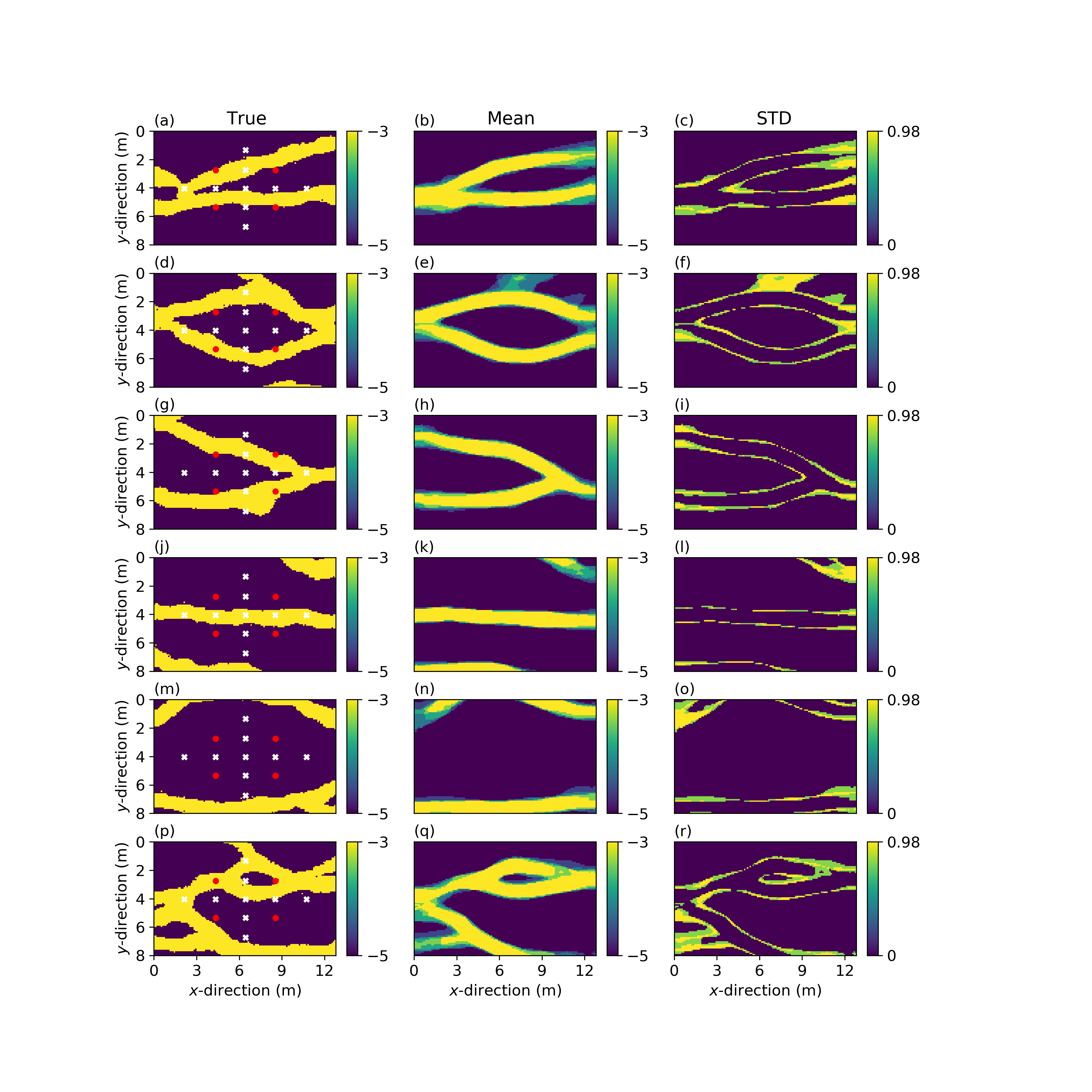}
	\caption{\small{Predictive uncertainty of inverse mapping using deep ensembles associated with the previously selected six true models for the case with transient subsurface pressure data and a binary channelized domain. The True label indicates the true model while the Mean and STD labels denote the mean predicted model and its associated standard deviation map calculated over an ensemble of five members. Table \ref{table10} lists the prediction quality statistics associated with the mean model and the five ensemble models.}} 
	\label{fig10}
\end{figure}

To assess predictive uncertainty using deep ensembles \citep[][]{Fort-Jastrzebski2019}, we trained the networks five times using random initialization and present the results obtain for the multi-Gaussian (Figure \ref{fig9} and Table \ref{table9}) and binary channelized hydraulic cases (Figure \ref{fig10} and Table \ref{table10}). These estimates should only be considered qualitatively as it is based on a small ensemble of five members and the resulting errors only refer to errors induced by training the neural networks. For instance, it is difficult to see a clear pattern for the multi-Gaussian case, except for an expected general tendency of larger uncertainties away from the measurement points. For the categorical case, the predictive uncertainty is as expected the largest at boundaries between the two categories \citep[c.f., Figure 3 in][]{Zahner2016} with the thickness of the uncertainty bands being the largest at the sides of the domain, that is, the furthest away from the measurement points. Moreover, all members of the considered ensembles capture the same main spatial patterns (Figures \ref{fig9} and \ref{fig10}) and show comparable performance (Tables \ref{table9} and \ref{table10}).

\begin{table}[hbt!]
	\caption{\small{Statistics of the ensemble-based uncertainty quantification results obtained for transient subsurface pressure data in a multi-Gaussian domain. The True and Mean models are those depicted in Figure \ref{fig8}. The mean model is obtained from the deep ensemble models \#1 to \#5. RMSE$_{\rm data}$ denotes the RMSE in data space, ${\rm l}_1$ refers to the ${\rm l}_1$-norm and SSIM to the structural similarity index. The ${\rm l}_1$-norm is calculated in terms of $\log_{10}K_s$ (-) while the SSIM is computed in the rescaled $\left[0,1\right]$ domain.}}
	\begin{adjustbox}{center}
		\begin{tabular}{cccc}%
			\hline
			True model & RMSE$_{\rm data}$ (m) & ${\rm l}_1$ (m) & SSIM (-)\\
			\hline
			(a) True & 0.010 & 0 & 1 \\
			(b) Mean & 0.032 & 1858 & 0.71\\
			(d) Model \#1 & 0.023 & 1837 & 0.73 \\
			(e) Model \#2 & 0.040 & 1965 & 0.68 \\
			(f) Model \#3 & 0.025 & 1949 & 0.67 \\
			(g) Model \#4 & 0.032 & 1876 & 0.73 \\
			(h) Model \#5 & 0.028 & 2073 & 0.67 \\
			
			(i) True & 0.010 & 0 & 1 \\
			(j) Mean & 0.068 & 1597 & 0.73\\
			(l) Model \#1 & 0.054 & 1559 & 0.76 \\
			(m) Model \#2 & 0.046 & 1779 & 0.68 \\
			(n) Model \#3 & 0.055 & 1513 & 0.72 \\
			(o) Model \#4 & 0.059 & 1660 & 0.71 \\
			(p) Model \#5 & 0.061 & 1922 & 0.64 \\
			
			(q) True & 0.010 & 0 & 1 \\
			(r) Mean & 0.016 & 1571 & 0.77\\
			(t) Model \#1 & 0.016 & 1525 & 0.75 \\
			(u) Model \#2 & 0.020 & 1572 & 0.75 \\
			(v) Model \#3 & 0.015 & 1795 & 0.72 \\
			(w) Model \#4 & 0.018 & 1492 & 0.77 \\
			(x) Model \#5 & 0.016 & 1752 & 0.74 \\
			
			\hline
		\end{tabular}
	\end{adjustbox}
	\label{table9}
\end{table}

\begin{table}[hbt!]
	\caption{\small{Statistics of the ensemble-based uncertainty quantification results obtained for transient subsurface pressure data in a binary channelized domain. The True and Mean models are those depicted in Figure \ref{fig9}. The mean model is obtained from the deep ensemble models \#1 to \#5. RMSE$_{\rm data}$ denotes the RMSE in data space, ${\rm l}_1$ refers to the ${\rm l}_1$-norm and SSIM to the structural similarity index. The ${\rm l}_1$ is calculated in terms of $\log_{10}K_s$ (-) while the SSIM is computed in the rescaled $\left[0,1\right]$ domain.}}
	\begin{adjustbox}{center}
		\begin{tabular}{cccc}%
			\hline
			True model & RMSE$_{\rm data}$ (m) & ${\rm l}_1$ (m) & SSIM (-)\\
			\hline
			(a) True & 0.010 & 0 & 1 \\
			(b) Mean & 0.577 & 2492 & 0.67\\
			(d) Model \#1 & 0.590 & 2594 & 0.67 \\
			(e) Model \#2 & 0.095 & 2908 & 0.66 \\
			(f) Model \#3 & 0.267 & 2254 & 0.71 \\
			(g) Model \#4 & 0.523 & 2912 & 0.64 \\
			(h) Model \#5 & 0.523 & 1782 & 0.73 \\
			
			(i) True & 0.010 & 0 & 1 \\
			(j) Mean & 0.449 & 2321 & 0.62\\
			(l) Model \#1 & 0.216 & 2246 & 0.65 \\
			(m) Model \#2 & 0.426 & 2558 & 0.63 \\
			(n) Model \#3 & 0.183 & 2486 & 0.63 \\
			(o) Model \#4 & 0.187 & 1988 & 0.67 \\
			(p) Model \#5 & 0.233 & 2356 & 0.63 \\
			
			(q) True & 0.010 & 0 & 1 \\
			(r) Mean & 0.377 & 3046 & 0.60\\
			(t) Model \#1 & 0.416 & 3334 & 0.60 \\
			(u) Model \#2 & 0.252 & 3422 & 0.58 \\
			(v) Model \#3 & 0.271 & 3226 & 0.59 \\
			(w) Model \#4 & 0.049 & 2520 & 0.64 \\
			(x) Model \#5 & 0.397 & 2728 & 0.63 \\
			\hline
		\end{tabular}
	\end{adjustbox}
	\label{table10}
\end{table}
\FloatBarrier

\subsection{Effect of the Size of the Training Set}
\label{size_tr}

\begin{figure}[hbt!]
	\noindent\hspace{-4cm}\includegraphics[width=60pc]{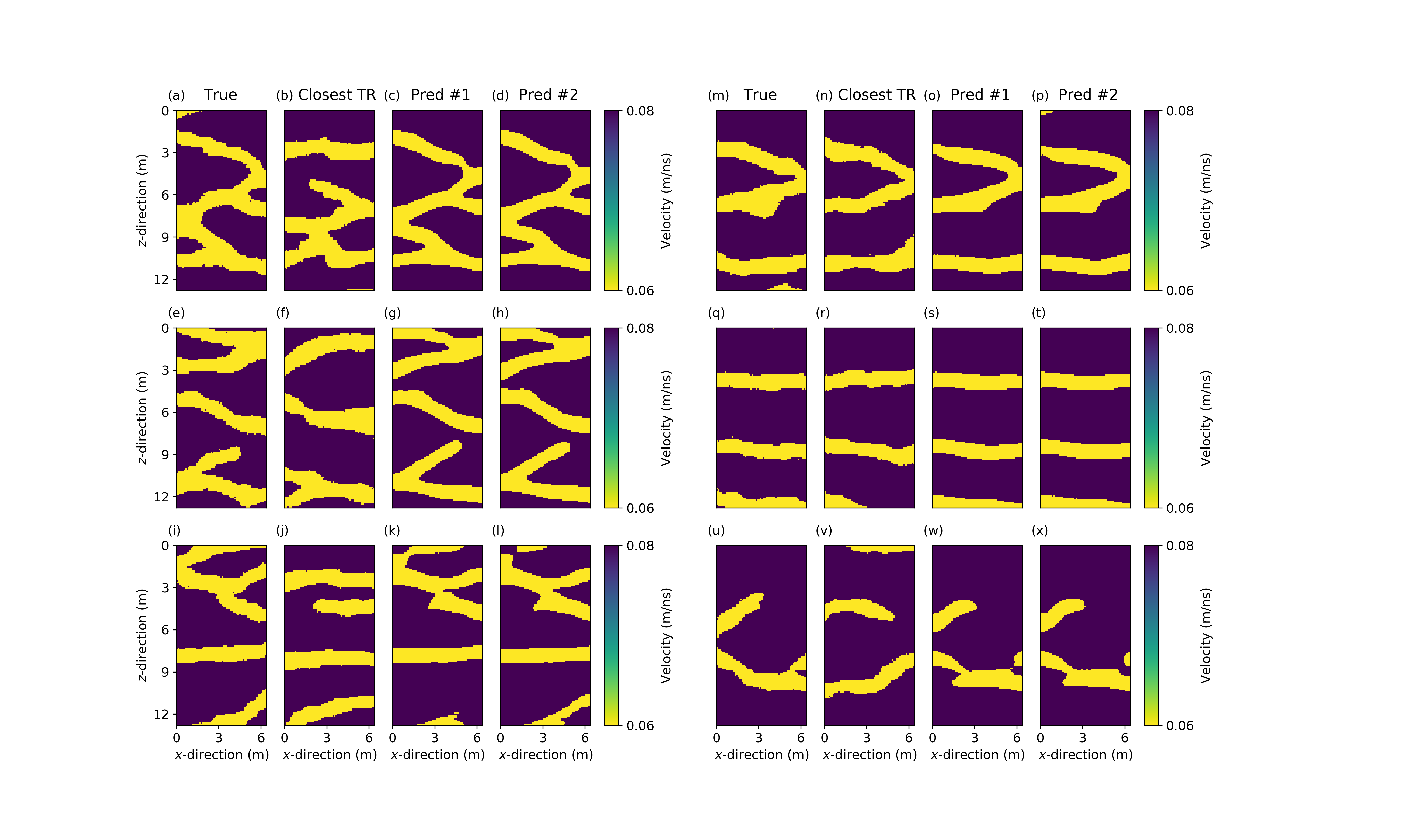}
	\caption{\small{Results for crosshole GPR data in a binary channelized domain when \textit{vec2pix} is trained with 5000 training examples only. (a-d) The true model is the most different test model from the set of 20,000 training models in data space: (a) true model, (b) closest training model in data space, (c, d) predicted models from two different noise realizations (Pred \#1 and Pred \#2). The same plotting style is adapted for cases (e-h) where the true model is the second most different model test model in data space, and (i-l), (m-p), (q-t) and (m-x) for four representative test models.}} 
	\label{fig11}
\end{figure}

\begin{figure}[hbt!]
	\noindent\hspace{-2cm}\includegraphics[width=50pc]{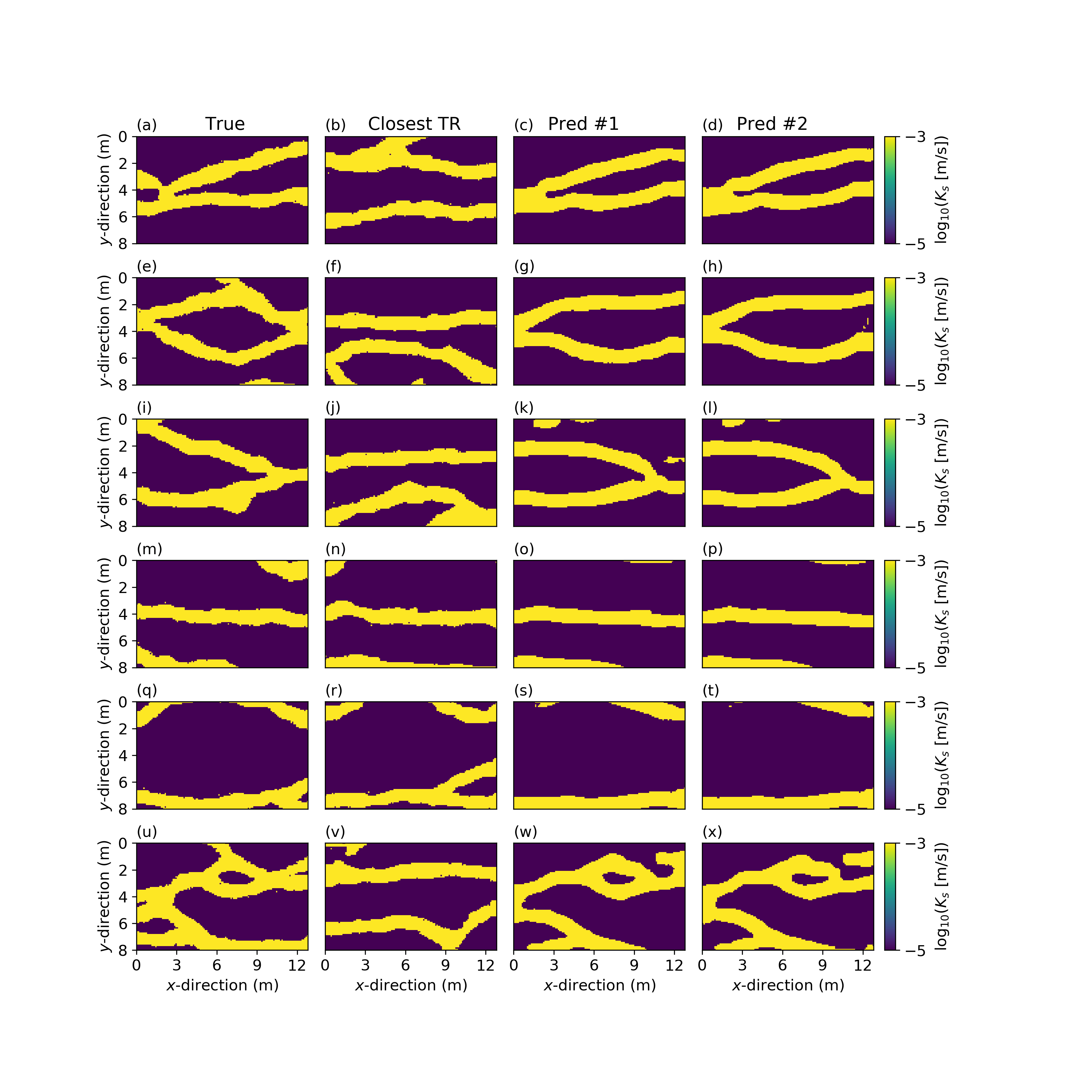}
	\caption{\small{Results for transient subsurface pressure data in a binary channelized domain when \textit{vec2pix} is trained with 5000 training examples only. (a-d) The true model is the most different test model from the set of 20,000 training models in data space: (a) true model, (b) closest training model in data space, (c, d) predicted models from two different noise realizations (Pred \#1 and Pred \#2). The same plotting style is adapted for cases (e-h) where the true model is the second most different model test model in data space, and (i-l), (m-p), (q-t) and (m-x) for four representative test models.}} 
	\label{fig12}
\end{figure}

The experiments described in sections \ref{results:case1} to \ref{results:case4} were repeated using 5000 and 10,000 training examples to train \textit{vec2pix}. The resulting peformance statistics for the same true models as those considered before are reported in Tables \ref{table2}, \ref{table4}, \ref{table6} and \ref{table8}. Compared to using a training set size of 20,000, reducing the size of the training set degrades performance only moderately. For instance, it remains a better strategy to train \textit{vec2pix} with 5000 training examples than to pick up the best model in the data space among 20,000 examples. Figures \ref{fig11} and \ref{fig12} depict models produced by \textit{vec2pix} when trained with 5000 training examples for the case of a binary channelized model domain. Even if the results are less good than the corresponding results in Figures \ref{fig6} and \ref{fig8} for 20,000 training examples, we still find that the produced models are visually close to the true models. Typical evolutions of the $\rm{l}_1$ loss during training are depicted in Figure \ref{fig2} for the three training set sizes and the case of transient flow within a binary channelized model domain (Case study 4, section \ref{results:case4}). It is observed that for every training set size the major $\rm{l}_1$ reduction occurs within the first 50 - 100 training epochs.

\FloatBarrier

\subsection{Pure Regression Versus Adversarial Learning}

Previous geoscientific applications of deep domain transfer based on the \textit{pix2pix} - \textit{cycleGAN} framework relied on adversarial training \citep{Goodfellow2014} of $G_{YX}$. We have done so herein too and, as described below, noticed no real added value of using adversarial learning. We tested with the Wasserstein generative adversarial network \citep[WGAN,][]{Arjovsky2017} training framework, in which the Wasserstein distance between the probability distributions of the true and generated data is minimized. Furthermore, we used the Wasserstein GAN with gradient penalty \citep[WGANGP,][]{Gulrajani2017} which has been shown to further stabilize training compared to the WGAN.

The full model now consists of the mapping function, $G_{YX}$ with an associated critic function, $D_{X}$:
	\begin{equation}
	\label{eq4}
	G_{YX} : \mathbb{R}^Y \rightarrow \mathbb{R}^X, D_{X} : \mathbb{R}^X \rightarrow \left[0,1\right].
	\end{equation}
The critic or discriminator, $D_{X}$, tries to distinguish between the true and predicted models, $\textbf{x}$ and $\tilde{\textbf{x}}$. For this experiment we used a fully convolutional  $D_{X}$ such as in \citet{Laloy2018}. At training time, $G_{YX}$ and $D_{X}$ are jointly learned using the sum of two losses: an adversarial loss and a reconstruction loss. The motivation for using an adversarial loss is to ensure that a realistically-looking $\tilde{\textbf{x}}$ model is predicted for any given $\textbf{y}$ vector, while as for pure regression training the reconstruction loss is required to enforce that each $\tilde{\textbf{x}}$ is in close agreement with its corresponding $\textbf{x}$.

The WGANGP objective function is given by
	\begin{equation}
	\label{eq5}
	\begin{split}
	\mathcal{L}_{\rm WGAN}\left(G_{YX},D_{X}, \textbf{y}, \textbf{x} \right) = \underset{\textbf{y}\sim p_{\textbf{y}}}{\mathbb{E}}\left[ D_{X}\left(G_{YX}\left(\textbf{y}\right)\right) \right] - \underset{\textbf{x}\sim p_{\textbf{x}}}{\mathbb{E}}\left[ D_{X}\left(\textbf{x}\right) \right] + \\ \lambda_{\rm GP} \underset{\hat{\textbf{x}}\sim p_{\hat{\textbf{x}}}}{\mathbb{E}}\left[\left(||\nabla_{\hat{\textbf{x}}} D_{X}\left(\hat{\textbf{x}}\right)||_2 - 1\right)^2\right],
	\end{split}
	\end{equation}

where $p_{\hat{\textbf{x}}}$ is sampling uniformly along straight lines between pairs of points sampled from the data distribution, $p_{\textbf{x}}$, and the generator distribution, $p_{\tilde{\textbf{x}}}$. This means that the $\hat{\textbf{x}}$ models are interpolations between the real, $\textbf{x}$, the and generated, $\tilde{\textbf{x}}$, models. The penalty coefficient, $\lambda_{\rm GP}$, is set to 10 \citep[][]{Gulrajani2017}.

Combining equations (\ref{eq2}) and (\ref{eq5}), the full objective function for training \textit{vec2pix} becomes
	\begin{equation}
	\label{eq6}
	\min_{G_{YX}} \max_{D_{X}}\left\{\mathcal{L}_{\rm WGAN}\left(G_{YX},D_{X}, \textbf{y}, \textbf{x} \right) + \lambda_{\rm rec}\mathcal{L}_{\rm rec}\left(G_{YX}, \textbf{y}, \textbf{x} \right)\right\},
	\end{equation}
where  $\lambda_{\rm rec}$ determines the relative importance of each objective. Extensive testing revealed that $\lambda_{\rm rec}$ needs to be set to a rather large value to get the most accurate reconstruction: $\lambda_{\rm rec} \geq 10^{5}$. Given the actual values taken by $\mathcal{L}_{\rm WGAN}\left(G_{YX},D_{X}, \textbf{y}, \textbf{x} \right)$ and $\lambda_{\rm rec}\mathcal{L}_{\rm rec}\left(G_{YX}, \textbf{y}, \textbf{x} \right)$, this means that the adversarial loss has virtually no influence on the total loss function, and therefore, adversarial training is not needed for the problems considered herein.

\section{Discussion}
\label{discussion}

We have introduced \textit{vec2pix}, a deep neural network for predicting 2D subsurface property fields from one-dimensional measurement data (e.g., time series). Our approach is illustrated using (1) synthetic crosshole first-arrival GPR travel times for recovering a 2D velocity field, and (2) time series of transient hydraulic heads to infer a 2D hydraulic conductivity field. For each problem, both a multi-Gaussian and a binary channelized subsurface domain with long-range connectivity are considered. Training \textit{vec2pix} is achieved using (at most) 20,000 training examples. For every considered case, our method is found to retrieve a model that is much closer to the true model than the closest training model in data space. Even if these recovered models generally look similar to the true models, the data RMSE obtained when forward simulating the \textit{vec2pix} models are higher than the prescribed noise level (that is, Gaussian white noise used to contaminate the true data). This is particularly true for our fourth case study that considers a transient pumping experiment within a channelized subsurface domain for which the relationship between model and simulated pressure data is highly nonlinear and, to some extent, not unique. If data fitting to the noise level is needed, we suggest that the solution derived by \textit{vec2pix} could be used as a starting point for a multiple-point statistics (MPS) based inversion such as sequential geostatistical resampling \citep[e.g.,][]{Mariethoz2010b}. The computational cost incurred by this additional MPS-based step will largely depend on the quality of the \textit{vec2pix}-derived solutions.

Uncertainty quantification based on deep ensembles obtained by training the network repeatedly with random initialization provides, at least, qualitatively-meaningful results. Although we only considered ensembles of five, we note that the uncertainty grows as expected with distance from the measurement points. For the binary channelized case, we reproduce similar patterns of uncertainty as found with completely different inversion approaches \citep[][]{Zahner2016}, with the uncertainty being the highest at channel boundaries. More work with larger deep ensembles is needed to understand to which extent the uncertainty quantification can be interpreted more quantitatively. We clarify that the uncertainty assessed herein is not directly based on the data misfit, that is, the produced ensemble models may not induce a data misfit that is in accordance with the underlying noise model. So a suggested above, for a pure Bayesian interpretation of uncertainty it is necessary to use the \textit{vec2pix} results as a starting point for a MCMC-based MPS inversion.

A training set of 20,000 examples was used as baseline in this study. Considering a larger training base would likely further improve prediction quality, but at the cost of a larger computational demand since obtaining one training sample requires one forward model evaluation. When reducing the training set to only 10,000 and 5000 training examples, we find that the resulting reduction in performance is rather moderate. This suggests that \textit{vec2pix} could be used with less training examples than 20,000 either for less complex subsurface heterogeneities than those considered in this work or if less accurate results are acceptable. In this trade-off between cost of building the training set and \textit{vec2pix} performance, the actual choice of the training size needs to be determined based on the test example and study objectives. Furthermore, we stress that parallel computing can be used to build the training set since this is to be done offline. The resulting speedup can thus be large if many parallel cores are available. Physics-informed deep networks \citep[e.g.,][]{Raissi2019} have been shown to drastically reduce the training needs when building proxy forward models. Including such concepts for inverse mapping could further reduce the training demand.

Prior studies relying on the \textit{pix2pix} - \textit{cycleGAN} framework \citep[][]{Mosser2018,Sun2018} used adversarial training \citep{Goodfellow2014}, of which the current standard is the Wasserstein generative adversarial network \citep[WGAN,][]{Arjovsky2017}. We evaluated this alternative herein, using the state-of-the-art Wasserstein GAN with gradient penalty \citep[WGANGP,][]{Gulrajani2017} method. Doing so, the total loss function used to train $G_{YX}$ becomes a weighted sum of two losses: the WGANGP loss (equation (\ref{eq5})) and the reconstruction loss (equation (\ref{eq2})). We observed that for the considered case studies, adversarial training is not needed. Indeed, to achieve the most accurate results the relative weight of the reconstruction loss compared to the WGANGP loss needs to be so large that the WGANGP loss has negligible influence on the total loss function.

As an alternative to the \textit{vec2pix} architecture, we tested for the flow problem if it is better to reshape the 1D input vector to 2D at the entry of the network instead of achieving this reshaping in the center of our ``diabolo"-like network (Figure \ref{fig1}). Since padding the input 2D matrix is necessary, we considered two padding options: zero-padding and replication padding. Our results consistently showed a 10\% reduction in performance based on the $\rm{l}_1$ norm.

In field applications, the measurements presented to \textit{vec2pix} will be contaminated with measurement errors. This is why we trained \textit{vec2pix} with noise-contaminated data. However, limited testing showed that noise-corrupting the training data or not does not lead to important differences at test time, when the test data are noise-contaminated. That said, we have used realistic, but low, noise levels to corrupt our data and the situation may change if larger noise values are prescribed. Furthermore, real-world applications will bring more complexities such as forward model errors and the degree of inadequacy of the prior geologic model (training image). This warrants further investigations with real data. 

Lastly, our approach requires a new training of \textit{vec2pix} for each measurement configuration (measurement locations and acquisition times). This limitation does not apply to GAN-based inversion \citep[][]{Laloy2018,Laloy2019} where a GAN is trained once for a given training image and inversions of different direct and indirect measurement datasets can be performed in the latent space of the GAN \citep[][]{Laloy2018,Mosser2020}. However, GAN-based inversion still has a substantial computational demand when done probabilistically \citep[][]{Laloy2018}, while the nonlinearity of the GAN transform may prevent deterministic gradient-based inversion \citep[][]{Laloy2019} from being effective.

\section{Conclusion}
\label{conclusion}
We introduce \textit{vec2pix}, a deep neural network for predicting categorical and continuous 2D subsurface property fields from one-dimensional measurement data (e.g., time series) and, thereby, offering an alternative approach to solve inverse problems. The method is illustrated using (1) synthetic first-arrival GPR travel times to infer a 2D velocity field, and (2) synthetic time series of transient hydraulic heads to retrieve a 2D hydraulic conductivity field. For each problem type, both a multi-Gaussian and a binary channelized subsurface domain with long-range connectivity are considered. Using a training set of 20,000 examples, our approach always recovers a 2D model that is much closer to the true model than the closest training model in the forward-simulated data space. Despite a moderate decrease in performance, this remains also true when using only 5000 training examples. The inferred models generally look visually similar to the true ones, but the data misfits obtained when forward simulating these models are generally larger than the noise level used to corrupt the true data. To assess uncertainty, we have used a small deep ensemble, implying that the network is trained multiple times with random initialization. Qualitatively-speaking, these uncertainties are in agreement with the expected uncertainty patterns. This work opens up new perspectives on how to use deep learning to infer subsurface models from indirect measurement data.

\section{Acknowledgements}
We are grateful for helpful suggestions offered by Arnaud Doucet (deep ensembles) and Shiran Levy (choice of training rates). We are also thankful for the comments offered by the three anonymous reviewers. Upon acceptance of the manuscript, the code will be made available at \url{https://github.com/elaloy/vec2pix}. Part of our code is inspired by the \textit{cycleGAN-and-pix2pix} code by Jun-Yan Zhu (\url{https://github.com/junyanz/pytorch-CycleGAN-and-pix2pix}).

\section{Appendix: Network details}
\label{appendix}
The $G_{YX}$ network is made of convolutions, transposed convolutions and a series of ``ResNet" residual blocks \citep{He2016}. We use 6 residual blocks for cases involving binary images (or models) and 9 residual blocks for cases involving continuous images. Our used activation functions are either rectified linear unit: ReLU ($\max_{\left(0,x\right)}$) or hyperbolic tangent: Tanh, and we use reflection padding in the first and last layers of $G_{YX}$. Let $c_{\rm 2d}7-s1-k_{\rm in}1-k_{\rm out}64-p0$ denote a 7 $\times$ 7 2D Convolution-InstanceNorm-ReLU layer with $k_{\rm in} = 1$ incoming channels (or filters), $k_{\rm out} = 64$ outgoing channels, stride 1 and zero padding. We call $co_{\rm 2d}7-s1-k_{\rm in}1-k_{\rm out}64-p0$ the same layer without normalization and with a Tanh activation function. Furthermore, $tc_{\rm 2d}$ signifies a 2D Transposed Convolution-InstanceNorm-ReLU, $op1$ means output padding of 1 and $R_{\rm 2d}-k512$ represents a residual block that contains two 3 $\times$ 3 2D convolutional layers with InstanceNorm and $k=512$ channels on both layers, and a ReLU activation function on the first layer. Lastly, $Re\left(z_r,z_c\right)$ and $Fla$ mean reshaping a vector into a $z_r \times z_c$ array and flattening an 2D array, respectively. From input to output layer, our generator is built as follows
\begin{itemize}
	\item $\left[c_{\rm 1d}7-s1-k_{\rm in}1-k_{\rm out}64-p0\right]$
	\item $\left[c_{\rm 1d}3-s2-k_{\rm in}64-k_{\rm out}128-p1\right]$
	\item $\left[c_{\rm 1d}3-s2-k_{\rm in}128-k_{\rm out}256-p1\right]$
	\item $\left[c_{\rm 1d}3-s2-k_{\rm in}256-k_{\rm out}512-p1\right]$
	\item $Re\left(z_r,z_c\right)$
	\item $N_{\rm res} \times \left[R_{\rm 2d}-k512\right]$
	\item $\left[tc_{\rm 2d}3-s2-k_{\rm in}512-k_{\rm out}256-p1-op1\right]$
	\item $\left[tc_{\rm 2d}3-s2-k_{\rm in}256-k_{\rm out}128-p1-op1\right]$
	\item $\left[tc_{\rm 2d}3-s2-k_{\rm in}128-k_{\rm out}64-p1-op1\right]$
	\item $\left[co_{\rm 2d}7-s1-k_{\rm in}64-k_{\rm out}1-p0\right]$	
\end{itemize}
where $\displaystyle z_r = \frac{X_r}{8}$ and $\displaystyle z_c = \frac{X_c}{8}$ with $X_r$ and $X_c$ the numbers of rows and columns of a model $\textbf{X}$, the incoming data vector \textbf{y} is padded with zeros such as its size matches $\frac{X_rX_c}{8}$, and $N_{\rm res}$ is the selected number of residual blocks (6 or 9, see above).

\bibliographystyle{unsrt}  


\end{document}